\documentclass[10pt, conference]{IEEEtran}
\IEEEoverridecommandlockouts
\usepackage{cite}
\usepackage{xspace}
\usepackage{makecell}
\usepackage{float}
\usepackage{amsmath,amssymb,amsfonts}
\usepackage{algorithmic}
\usepackage{enumitem}
\usepackage{graphicx}
\usepackage{textcomp}
\usepackage{xcolor}
\usepackage{booktabs}
\usepackage{multirow}
\usepackage{caption}
\usepackage{multicol}
\usepackage{tablefootnote}
\usepackage{threeparttable}
\usepackage[hidelinks]{hyperref}

\def\BibTeX{{\rm B\kern-.05em{\sc i\kern-.025em b}\kern-.08em
    T\kern-.1667em\lower.7ex\hbox{E}\kern-.125emX}}

\begin{document}

\newcommand{\CodeIn}[1]{{\small \texttt{#1}}}
\newcommand{\mypara}[1]{\vspace{.03in}\noindent \textbf{#1}}
\newcommand{\Contrib}{$\star$\xspace}
\newcommand{\STAB}[1]{\begin{tabular}{@{}c@{}}#1\end{tabular}}
\newlist{researchquestions}{enumerate}{1}
\setlist[researchquestions]{label*=\textbf{RQ\arabic*}}

\newcommand{\alpharepair}{AlphaRepair\xspace} %
\newcommand{\codebert}{CodeBERT\xspace}

\newcommand{\inputseqfull}{X = \{x_1, x_2, ... x_n\}\xspace}
\newcommand{\inputseq}{X\xspace}
\newcommand{\inputseqone}{x_i\xspace}
\newcommand{\maskseq}{X_{masked}\xspace}
\newcommand{\maskcode}{C_{masked}\xspace}
\newcommand{\maskword}{W_{masked}\xspace}
\newcommand{\predictor}{P\xspace}
\newcommand{\lossmlm}{\mathcal{L}_{\mlm}\xspace}
\newcommand{\lossmlmbi}{\mathcal{L}_{bimodal\_\mlm}\xspace}
\newcommand{\blinelength}{L\xspace}
\newcommand{\gmasklines}{G\xspace}
\newcommand{\tjs}{temp joint score\xspace}
\newcommand{\js}{joint score\xspace}
\newcommand{\beamwidth}{N\xspace}
\newcommand{\tokenslength}{n\xspace}
\newcommand{\tokensgeneratedlength}{p\xspace}
\newcommand{\codebertpredictor}{C\xspace}
\newcommand{\codebertpredictortemp}{C^*\xspace}
\newcommand{\generatedseqfull}{T = \{t_1, ... t_p, MASK_{p+1}, ... MASK_{n}\}\xspace}
\newcommand{\generatedseq}{T}
\newcommand{\generatedseqone}{t}
\newcommand{\generatedseqi}{t_i}
\newcommand{\generatedseqp}{t_p}
\newcommand{\generatedseqprev}{\{t_1, ... t_{p-1}\}\xspace}
\newcommand{\generatedseqoneless}{\{t_1, ... t_{p-1}, t_p, MASK_{p+1}, ... MASK_{n}\}\xspace}
\newcommand{\generatedseqafter}{\{MASK_{p+1}, ... MASK_{n}\}\space}
\newcommand{\fullgeneratedseq}{T = \{t_1, t_2, ... t_n\}\xspace}
\newcommand{\rewardrepair}{RewardRepair\xspace}
\newcommand{\recoder}{Recoder\xspace}
\newcommand{\deepdebug}{DeepDebug\xspace}
\newcommand{\cure}{CURE\xspace}
\newcommand{\coconut}{CoCoNuT\xspace}
\newcommand{\dlfix}{DLFix\xspace}
\newcommand{\sequencer}{SequenceR\xspace}
\newcommand{\tbar}{TBar\xspace}
\newcommand{\prapr}{PraPR\xspace}
\newcommand{\avatar}{AVATAR\xspace}
\newcommand{\simfix}{SimFix\xspace}
\newcommand{\fixminer}{FixMiner\xspace}
\newcommand{\capgen}{CapGen\xspace}
\newcommand{\jaid}{JAID\xspace}
\newcommand{\sketchfix}{SketchFix\xspace}
\newcommand{\nopol}{NOPOL\xspace}
\newcommand{\jgenprog}{jGenProg\xspace}
\newcommand{\jmutrepair}{jMutRepair\xspace}
\newcommand{\jkali}{jKali\xspace}
\newcommand{\genprog}{GenProg\xspace}
\newcommand{\angelix}{Angelix\xspace}
\newcommand{\spr}{SPR\xspace}
\newcommand{\aande}{AE\xspace}
\newcommand\arja{ARJA\xspace}
\newcommand\kalia{Kali-A\xspace}
\newcommand\card{Cardumen\xspace}
\newcommand\npefix{NPEFix\xspace}
\newcommand\gproga{GenProg-A\xspace}
\newcommand\rsrepair{RSRepair-A\xspace}
\newcommand\dynamoth{DynaMoth\xspace}

\newcommand{\codex}{Codex\xspace}
\newcommand{\gptj}{GPT-J\xspace}
\newcommand{\gptneo}{GPT-Neo\xspace}
\newcommand{\gptneox}{GPT-NeoX\xspace}
\newcommand{\incoder}{\textsc{InCoder}\xspace}
\newcommand{\codet}{CodeT5\xspace}

\newcommand{\plm}{PLM\xspace}
\newcommand{\plmfull}{Large Pre-Trained Language Model\xspace}
\newcommand{\clmfull}{Generative Model\xspace}
\newcommand{\clm}{generative model\xspace}
\newcommand{\apr}{APR\xspace}
\newcommand{\aprfull}{Automated Program Repair\xspace}
\newcommand{\nmt}{NMT\xspace}
\newcommand{\nmtfull}{Neural Machine Translation\xspace}
\newcommand{\nlp}{NLP\xspace}
\newcommand{\nlpfull}{Natural Language Processing\xspace}
\newcommand{\mlm}{infilling model\xspace}
\newcommand{\mlmfull}{Infilling Model\xspace}
\newcommand{\rtd}{RTD\xspace}
\newcommand{\rtdfull}{Replaced Token Detection\xspace}

\newcommand{\dfj}{Defects4J\xspace}
\newcommand{\quix}{QuixBugs\xspace}
\newcommand{\bears}{Bears\xspace}
\newcommand{\bip}{BugsInPy\xspace}
\newcommand{\manybugs}{ManyBugs\xspace}

\newcommand{\learning}{learning-based\xspace}
\newcommand{\Learning}{Learning-based\xspace} %
\newcommand{\template}{template-based\xspace}
\newcommand{\heuristic}{heuristic-based\xspace}
\newcommand{\constraint}{constraint-based\xspace}

\newcommand{\totaltotal}{1094\xspace}
\newcommand{\totalsingle}{572\xspace}
\newcommand{\totalhunk}{413\xspace}
\newcommand{\totalline}{246\xspace}
\newcommand{\bearstotal}{251\xspace}
\newcommand{\bearssingle}{51\xspace}
\newcommand{\bearshunk}{37\xspace}
\newcommand{\bearsline}{16\xspace}

\newcommand{\bugsinpytotal}{501\xspace}
\newcommand{\bugsinpysingle}{126\xspace}
\newcommand{\bugsinpyhunk}{89\xspace}
\newcommand{\bugsinpyline}{35\xspace}

\newcommand{\quixjavatotal}{40\xspace}
\newcommand{\quixjavasingle}{40\xspace}
\newcommand{\quixjavahunk}{37\xspace}
\newcommand{\quixjavaline}{36\xspace}

\newcommand{\quixpythontotal}{40\xspace}
\newcommand{\quixpythonsingle}{40\xspace}
\newcommand{\quixpythonhunk}{40\xspace}
\newcommand{\quixpythonline}{40\xspace}

\newcommand{\manytotal}{185\xspace}
\newcommand{\manysingle}{39\xspace}
\newcommand{\manyhunk}{23\xspace}
\newcommand{\manyline}{12\xspace}
\newcommand{\manyrepro}{91\xspace}

\newcommand{\dfjonetotal}{391\xspace}
\newcommand{\dfjonesingle}{255\xspace}
\newcommand{\dfjonehunk}{154\xspace}
\newcommand{\dfjoneline}{80\xspace}

\newcommand{\dfjtwototal}{438\xspace}
\newcommand{\dfjtwosingle}{228\xspace}
\newcommand{\dfjtwohunk}{159\xspace}
\newcommand{\dfjtwoline}{78\xspace}

\captionsetup[figure]{font=bf,skip=2pt}%
\captionsetup[table]{font=bf,skip=2pt}%
\newcommand{\distance}{5pt}
\setlength{\textfloatsep}{1pt}%
\setlength{\floatsep}{\distance}%
\setlength{\intextsep}{\distance}%
\setlength{\dbltextfloatsep}{\distance} %
\setlength{\dblfloatsep}{\distance} %

\newcommand{\minus}{\scalebox{0.75}[1.0]{$-$}}

\title{Practical Program Repair in the Era of Large Pre-trained Language Models} %

\author{\IEEEauthorblockN{Chunqiu Steven Xia}
\IEEEauthorblockA{University of Illinois at\\Urbana-Champaign\\
chunqiu2@illinois.edu}
\and
\IEEEauthorblockN{Yuxiang Wei}
\IEEEauthorblockA{University of Illinois at\\ Urbana-Champaign\\
ywei40@illinois.edu}
\and
\IEEEauthorblockN{Lingming Zhang}
\IEEEauthorblockA{University of Illinois at\\ Urbana-Champaign\\
lingming@illinois.edu}}

\maketitle

\begin{abstract}
    \aprfull (\apr) aims to help developers automatically patch software bugs. However, current state-of-the-art traditional and \learning \apr techniques face the problem of limited patch variety, failing to fix complicated bugs. This is mainly due to the reliance on bug-fixing datasets to craft fix templates (traditional) or directly predict potential patches (\learning). \plmfull{s} (\plm{s}), trained using billions of text/code tokens, can potentially help avoid this issue. Very recently, researchers have directly leveraged \plm{s} for \apr without relying on any bug-fixing datasets. Meanwhile, such existing work either failed to include state-of-the-art \plm{s} or was not evaluated on realistic datasets. Thus, the true power of modern \plm{s} on the important \apr problem is yet to be revealed.  
    
    In this work, we perform the first extensive study on directly applying \plm{s} for \apr. We select 9 recent state-of-the-art \plm{s}, including both generative and infilling models, ranging from 125M to 20B in size. We designed 3 different repair settings to evaluate the different ways we can use \plm{s} to generate patches: 1) generate the entire patch function, 2) fill in a chunk of code given the prefix and suffix 3) output a single line fix. We apply the \plm{s} under these repair settings on 5 datasets across 3 different languages and compare different \plm{s} in the number of bugs fixed, generation speed and compilation rate. We also compare the \plm{s} against recent state-of-the-art \apr tools. Our study demonstrates that directly applying state-of-the-art \plm{s} can already substantially outperform all existing \apr techniques on all our datasets. Among the studied \plm{s}, the scaling effect exists for \apr where larger models tend to achieve better performance. Also, we show for the first time that suffix code after the buggy line (adopted in infilling-style \apr) is important in not only generating more fixes but more patches with higher compilation rate. Besides patch generation, the \plm{s} consider correct patches to be more \emph{natural} than other ones, and can even be leveraged for effective patch ranking or patch correctness checking. Lastly, we show that \plm-based \apr can be further substantially boosted via: 1) increasing the sample size, and 2) incorporating fix template information.%
\end{abstract}

\section{Introduction}

As software programs and systems become more and more ubiquitous in everyday life, so do software bugs. Due to the wide-ranging adoption of software systems in fields from healthcare~\cite{healthcare} to transportation~\cite{transportation}, these bugs can potentially cause dangerous safety issues~\cite{bug_safety} and financial losses~\cite{bug_loss}. As such, developers often need to spend a significant amount of time and effort to fix software bugs~\cite{debuggingtime}. In order to help developers reduce this manual effort, \aprfull (\apr) tools have been built to automatically generate potential patches given the original buggy program~\cite{gazzola2019aprsurvey}.

Among traditional \apr techniques~\cite{legoues2012genprog, le2016hdrepair, wen2018capgen, mechtaev2016angelix, le2017s3, demacro2014nopol, hua2018sketchfix, martinez2016astor, koyuncu2020fixminder, liu2019avatar}, template-based \apr has been widely recognized as the state of the art~\cite{liu2019tbar, ghanbari2019prapr}. These techniques leverage fix templates, often designed by human experts, to fix specific types of bugs in the source code. As a result, these \apr tools are constrained by the underlying fix templates in the types of bugs that can be fixed. To combat this, researchers have proposed \learning \apr tools~\cite{zhu2021recoder, jiang2021cure, ye2022rewardrepair, lutellier2020coconut}, which typically model program repair as a \nmtfull (\nmt) problem~\cite{sutskever2014mt}, where the goal is to translate a buggy program into a fixed program. The core component of these \learning \apr tools is an encoder and decoder pair, where the model aims to capture the buggy context via the encoder and then autoregressively generate the patch using the decoder. As such, these learning-based \apr tools require supervised training datasets containing pairs of buggy and patched code, usually obtained by mining historical bug fixes from open-source repositories. While \learning \apr tools have shown improvements in both the number and variety of bugs that can be fixed~\cite{zhu2021recoder, jiang2021cure}, they are still restricted by their training data which only contain a limited amount of bug fix types and may not generalize to unseen bug types~\cite{xia2022alpharepair}. 

Recent developments in building \plmfull{s} (\plm{s}) offer an alternative solution that can be applied for program repair without relying on historical bug fixes. While \plm{s} are usually general-purpose tools for \nlp tasks (e.g., GPT3~\cite{brown2020gpt3}), they have also been {used for} programming languages by finetuning on code (e.g., Codex~\cite{chen2021codex}). Unlike the specifically designed \learning \apr models, \plm{s} are trained in an unsupervised fashion using up to billions of text/code tokens and can be used in a variety of code tasks. Recently, \alpharepair~\cite{xia2022alpharepair} proposes to leverage \codebert~\cite{feng2020codebert}, a large code model pre-trained on millions of code snippets, directly for \apr. The key insight from \alpharepair is instead of learning transformations to go from buggy code to fixed code, we can directly use the model to predict what the correct code should look like given its surrounding context (including both prefix and suffix), i.e., \emph{infilling-style} \apr. Using this idea, \alpharepair demonstrated state-of-the-art repair results without finetuning on bug fixing dataset. While \alpharepair has shown improvements over previous \learning \apr, the model (125M parameters) it uses is far smaller than the current state-of-the-art \plm{s} (Codex: 12B parameters and GPT-3: 175B parameters). Beside \alpharepair, researchers have also directly leveraged \codex for \emph{generative} \apr~\cite{kolak2022patch,prenner2021codexws}, i.e., generating the fixes based on the context before bugs (i.e., prefix only). However, these studies mostly focus on Codex and are only evaluated on a small dataset with 40 bugs on simple programming tasks.

Current state-of-the-art \plm{s} such as \codex~\cite{chen2021codex} and \incoder~\cite{fried2022incoder} have also included evaluation for code related tasks such as code completion, docstring generation and variable/type prediction. However, these evaluations still mainly focus on \nlp metrics such as BLEU score~\cite{papineni2002bleu} which do not accurately measure the functional or semantic correctness of the generated code. Furthermore, the datasets consist of hand-curated code problems{ which} do not accurately reflect the type of projects developers work on in the real world.

\mypara{Our Work.} We present the first extensive evaluation of recent \plm{s} for fixing real-world projects. We designed 3 different \apr experimental settings: 1) complete function generation 2) correct code infilling and 3) single line generation to showcase the different ways \plm{s} can be applied for \apr. In our study, we include both popular types of \plm architectures (generative and \mlm{s}) to show the advantages and flaws of using each type for \apr. We include models with a wide range of different parameter sizes, spanning from 125 million to 20 billion. We evaluate not only the improvement in repair effectiveness but also the trade-off with respect to speed when increasing the model size. In total, we use 5 different repair datasets containing real open-source bugs and developer written tests across 3 programming languages to evaluate \apr under realistic settings. Compared with existing applications of \plm{s} for \apr~\cite{xia2022alpharepair, kolak2022patch, prenner2021codexws}, our study is the first to include state-of-the-art \plm{s} for both infilling-style and generative \apr on various datasets and programming languages. To summarize, this paper makes the following contributions.

\begin{itemize}[leftmargin=*]
    \item[\Contrib] \textbf{Dimension.} This paper bridges the gap between the recent advances in \plm{s} and a crucial software engineering problem -- \apr. This paper not only demonstrates the potential and future for directly leveraging \plm{s} for solving the important \apr problem, but also provides a realistic evaluation scenario for the recent \plm{s}, which were mainly evaluated on simple/synthetic coding problems rather than real-world systems as studied in the \apr area.%
    \item[\Contrib] \textbf{Study.} We conduct extensive evaluations using 9 different recent \plm{s} on 5 different repair datasets across 3 different programming languages (Java, Python, and C). We compare the \plm{s} against each other using the 3 repair settings we designed. Using the popular repair datasets, we further compare the \plm{s} with state-of-the-art \apr tools.%
    
    \item[\Contrib] \textbf{Practical Guidelines.} Our study shows for the first time that directly applying state-of-the-art \plm{s} can already substantially outperform all existing \apr tools on the widely studied \dfj 1.2 dataset (and other ones), e.g., Codex can fix 32 more bugs than the existing best \apr technique. Among the studied \plm{s}, the scaling effect exists for \apr where larger models tend to deliver stronger \apr results. Also, we show for the first time that suffix code after the buggy line (adopted in infilling-style \apr) is important in not only generating more fixes but more patches with higher compilation rate. Besides patch generation, the \plm{s} consider correct patches to be more \emph{natural} than other ones, and can even be used for effective patch ranking or correctness checking. Lastly, we show that \plm-based \apr can be further substantially improved via: 1) increasing the sample size, and 2) incorporating fix template information.

\end{itemize}

\section{Background and Related Work}

\subsection{\plmfull}

\plmfull{s} (\plm{s}) have become ubiquitous in the domain of \nlp, achieving impressive performance in many tasks such as machine translation~\cite{sutskever2014mt}, text summarization~\cite{liu2019finetune} and classification~\cite{yang2020xlnet}. \plm{s} follow the Transformer architecture~\cite{vaswani2017attention} -- an encoder to capture input representation and a decoder to generate output tokens. These \plm{s} are first pre-trained in an unsupervised manner, on large amounts of text data and then finetuned for downstream tasks. However, certain tasks may not have an abundance of finetuned data available. As such, researchers have evaluated the ability for \plm{s} to perform on downstream tasks without finetuning. This is achieved via prompt engineering~\cite{reynolds2021prompt} -- providing the model with natural language descriptions and demonstrations of the task it is trying to solve before giving the model the target input. This works by leveraging the general-purpose setup of \plm{s} where the unsupervised pretraining dataset already encompasses many domains of problems/tasks. Using this idea and the exponential growth in \plm size~\cite{kaplan2020scaling}, impressive performance in many tasks can be achieved even without any finetuning~\cite{brown2020gpt3}.

\plm{s} can be classified into encoder-only, decoder-only and encoder-decoder models based on their architectures. Encoder-only models (such as BERT~\cite{devlin2018bert}) contain only the \emph{encoder} component of a Transformer. They are typically designed to learn data representations and are trained using the Masked Language Modeling (MLM) objective -- a small percentage (e.g., 15\%) of tokens in the training data will be replaced by masked tokens, and then the models are trained to predict the original values of the masked tokens based on the bidirectional contexts. Decoder-only models (such as GPT-3~\cite{brown2020gpt3} and \gptneo~\cite{gpt-neo}) are large generative models that use the \emph{decoder} to predict the next token output given all previous tokens (i.e., left context or prefix only). To combine the usage of both encoder and decoder, encoder-decoder models (such as T5~\cite{raffel2020t5} and BART~\cite{lewis2019bart}) have also been proposed for sequence-to-sequence tasks where the training objective aims to recover the correct output sequence given the original input (e.g., corrupted to uncorrupted). One such training objective is span prediction tasks, where random spans (multiple tokens) are replaced with artificial span tokens and the model is tasked with recovering the original tokens. For inferencing, one can use the encoder-decoder models to infill text by also adding the artificial span token in place. Recently, researchers have also combined MLM with generative models to perform both bidirectional and autoregressive text generation or infilling~\cite{aghajanyan2022cm3}. In our \apr scenario, all types of \plm{s} can potentially be leveraged for generative or infilling-style \apr, and we select 9 state-of-the-art \plm{s} for our study (detailed in Section~\ref{sec:models}).

\subsection{\aprfull}

\aprfull (\apr) tools are used to generate patched code given the original code and the corresponding buggy location. Each patch generated by the \apr tool is validated against the test suite. \textit{Plausible patches} are ones which pass the entire suite. \textit{Correct patches} are plausible patches which correctly fix the underlying bug. 

Traditional \apr tools can be classified as heuristic-based~\cite{legoues2012genprog, le2016hdrepair, wen2018capgen}, constraint-based~\cite{mechtaev2016angelix, le2017s3, demacro2014nopol} and \template~\cite{hua2018sketchfix, martinez2016astor, koyuncu2020fixminder, liu2019tbar, liu2019avatar}. Traditionally, \template \apr tools achieve the best performance, where each template is handcrafted by human experts designed to provide a fix for a specific type of bug. However, these \template \apr tools can only fix the bug types that are part of the templates. As a result, researchers employed \learning \apr tools to generate more expressive patches. Learning-based \apr tools such as \recoder~\cite{zhu2021recoder}, \rewardrepair~\cite{ye2022rewardrepair}, and \cure~\cite{jiang2021cure} are based on \nmt techniques~\cite{sutskever2014mt} which require specific bug fixing data to train the \nmt model to generate a fix line given the buggy line. Due to this reliance on the bug-fixing data, these \learning tools are still limited in terms of the type of fixes it can apply. Recent work of \alpharepair~\cite{xia2022alpharepair} addresses this by performing \apr under a zero-shot setting by directly using the \codebert model for repair. \alpharepair fills the original buggy line with masked tokens and uses \codebert to replace the masked tokens with correct code tokens to generate repair, i.e., \emph{infilling-style} (also called \emph{cloze-style}) \apr. While \alpharepair is able to achieve state-of-the-art results, \codebert is considerably smaller than the newest \plm{s}. Additionally, \alpharepair is only tested on a single setting where the correct location of the buggy line is known. 

Recent work~\cite{prenner2021codexws, kolak2022patch} has also looked into directly applying \plm{s} for \apr. Prenner et al.~\cite{prenner2021codexws} conducted a small-scale evaluation for the \codex model on a simple dataset containing both Java and Python versions of buggy algorithm implementations. \codex is given the buggy function and by using prompt engineering, are then asked to generate a complete fixed function. The results show that \codex is competitive with state-of-the-art \learning \apr tools in Python but worse in Java. In contrast, we show that by using our practical repair settings, \plm{s} are able to outperform state-of-the-art \apr tools on both Java and Python. Kolak et al.~\cite{kolak2022patch} also used \codex along with 2 smaller \plm{s} and evaluated their ability to generate the correct patch line when given the code prefix on the same dataset as the previous work~\cite{prenner2021codexws}. The evaluation demonstrated the scaling effect of \plm{s} where the repair results can be improved by using larger models. Interestingly, the study leverages sum entropy for patch ranking while \alpharepair leverages mean entropy (i.e., both favors more \emph{natural}~\cite{hindle2012softwarenatural} patches). Thus, we also perform a study of leveraging various recent \plm{s} for computing both entropies for patch ranking on real-world systems.%

Overall, the 2 prior studies~\cite{prenner2021codexws, kolak2022patch} are done on a small dataset with synthetic bugs using only a small number of \plm{s}. Moreover, the input and repair setting being used in the studies are also limited, e.g., only considered \emph{generative} \apr. In this paper, we present an extensive study of applying various state-of-the-art \plm{s} for both infilling-style and generative \apr on diverse repair datasets across programming languages.

\section{Approach}

In this section we describe the \plm{s} selected for evaluation and introduce 3 different \apr generation settings we use to evaluate each \plm. These settings are designed to showcase the different practical ways we can directly use \plm{s} for \apr and highlight advantages and differences of the studied \plm types. Also, we detail the patch ranking strategy of using entropy to prioritize patches that are more likely to be correct.

\subsection{Models} 
\label{sec:models}
\begin{table}
    \caption{Evaluation \plm overview}
    \centering
    \label{tab:model_stats}
    \begin{tabular}{@{}lrcc@{}}
    \toprule
    \textbf{Model} & \textbf{\#Parameters} & \textbf{Training Dataset} & \textbf{Type} \\
    \midrule
    \gptneo & 125M/1.3B/2.7B & The Pile & Generative \\
    \gptj & 6.7B & The Pile  & Generative \\
    \gptneox & 20B & The Pile  & Generative \\
    \addlinespace[1pt]
    \codex & 12B & N.R.  & \makecell{Generative \\ \& Infilling}\\
    \codet & 220M & \makecell{CodeSearchNet \\ \& BigQuery} & Infilling \\
    \addlinespace[2pt]
    \incoder & 1.3B/6.7B & N.R.  & Infilling \\
    \bottomrule
    \end{tabular}

\end{table}

We begin by describing the different \plm{s} we use for evaluation. Our selection process starts with the list of popular models hosted on the Hugging Face~\cite{HuggingFaceWebPage} -- an open-source platform to host and deploy large models. We sort the list of models based on popularity (\#downloads this month) and select the \plm{s} which contain code as training data. Furthermore, we also pick models from different organizations and types (described below) to obtain a diverse set of models. Along with the open-source models, we also use the closed-source \codex model~\cite{chen2021codex} (accessible only via API) since it has shown to achieve impressive performance on code related tasks. In total, we use 9 different \plm{s} for our experiment. 

Our chosen \plm{s} range from 125M to 20B in parameter size. Table~\ref{tab:model_stats} presents the \plm overview. Column \textbf{Model} is the model name, \textbf{\#Parameters} presents the number of model parameters, \textbf{Training Dataset} indicates the dataset used for pre-training (N.R. is not released), and \textbf{Type} refers to the type of \apr the model can perform (infilling or generative).

\subsubsection{\clmfull{s}} 
\begin{itemize}[leftmargin=*]
    \item \textbf{\gptneo~\cite{gpt-neo}, \gptj~\cite{gpt-j}, \gptneox~\cite{gpt-neox-20b}} All three models are open-source implementations of the GPT-3 transformer architecture~\cite{brown2020gpt3}. In our experiments, we use \gptneo models with 125M, 1.3B and 2.7B parameters. \gptj and \gptneox are even larger models with 6.7B and 20B parameters. These models were trained on The Pile~\cite{gao2020pile}, an 800GB dataset combining 22 diverse text-based datasets with 7.6\% containing open-source Github code. 
    \item \textbf{\codex~\cite{chen2021codex}} A 12B parameter GPT-3 based model designed for code generation. Codex is initialized with GPT-3 weights trained on natural language corpus and then finetuned on a large corpus of 159GB code files. %
\end{itemize}

\subsubsection{\mlmfull{s}}
\begin{itemize}[leftmargin=*]
    \item \textbf{\codet~\cite{wang2021codet5}} A 220M parameter model based on T5~\cite{raffel2020t5} architecture designed for code related tasks. \codet is trained using span prediction objective on 8.35 million functions across 8 different programming languages by combining CodeSearchNet~\cite{husain2020codesearchnet} with C/C\# dataset from BigQuery~\cite{BigQuery}. 
    \item \textbf{\incoder~\cite{fried2022incoder}} A model designed for code infilling by adopting a causal masking objective~\cite{aghajanyan2022cm3}. \incoder is trained on both open-source Github/GitLab code (159 GB) and StackOverFlow questions and answers (57 GB). We use both the 1.3B and 6.7B parameter version. 
    \item \textbf{\codex} In addition to using \codex as a \clm, we use the recently added suffix feature~\cite{CodexSuffixAPI} to perform code infilling. Since \codex is not open-sourced, we do not know how the model performs the infilling. 
\end{itemize}

\subsection{\plm-based Patch Generation}
\label{sec:pipeline}

In our study, we designed three settings for \apr:

\noindent \textbf{1)\makebox[3pt]{}Complete function generation} -- the input is a buggy function and the goal is to output the patched function.

\noindent \textbf{2)\makebox[3pt]{}Correct code infilling} -- the buggy location is known and the goal is to generate the correct replacement code given the prefix and suffix of the buggy function.

\noindent \textbf{3)\makebox[3pt]{}Single line generation} -- the bug location is provided and the bug is fixed by a single line change. Single line generation uses a subset of bugs in correct code infilling. We separate this case since many fault-localization methods provides a ranking in the granularity of individual code lines~\cite{abreu2007ochiai, zhang2013injecting}. More importantly, both infilling and generative \plm{s} can be applied for this setting, enabling direct comparison of the two. We now describe the different inputs for each setting.

\begin{figure}
    \includegraphics[width=0.85\linewidth]{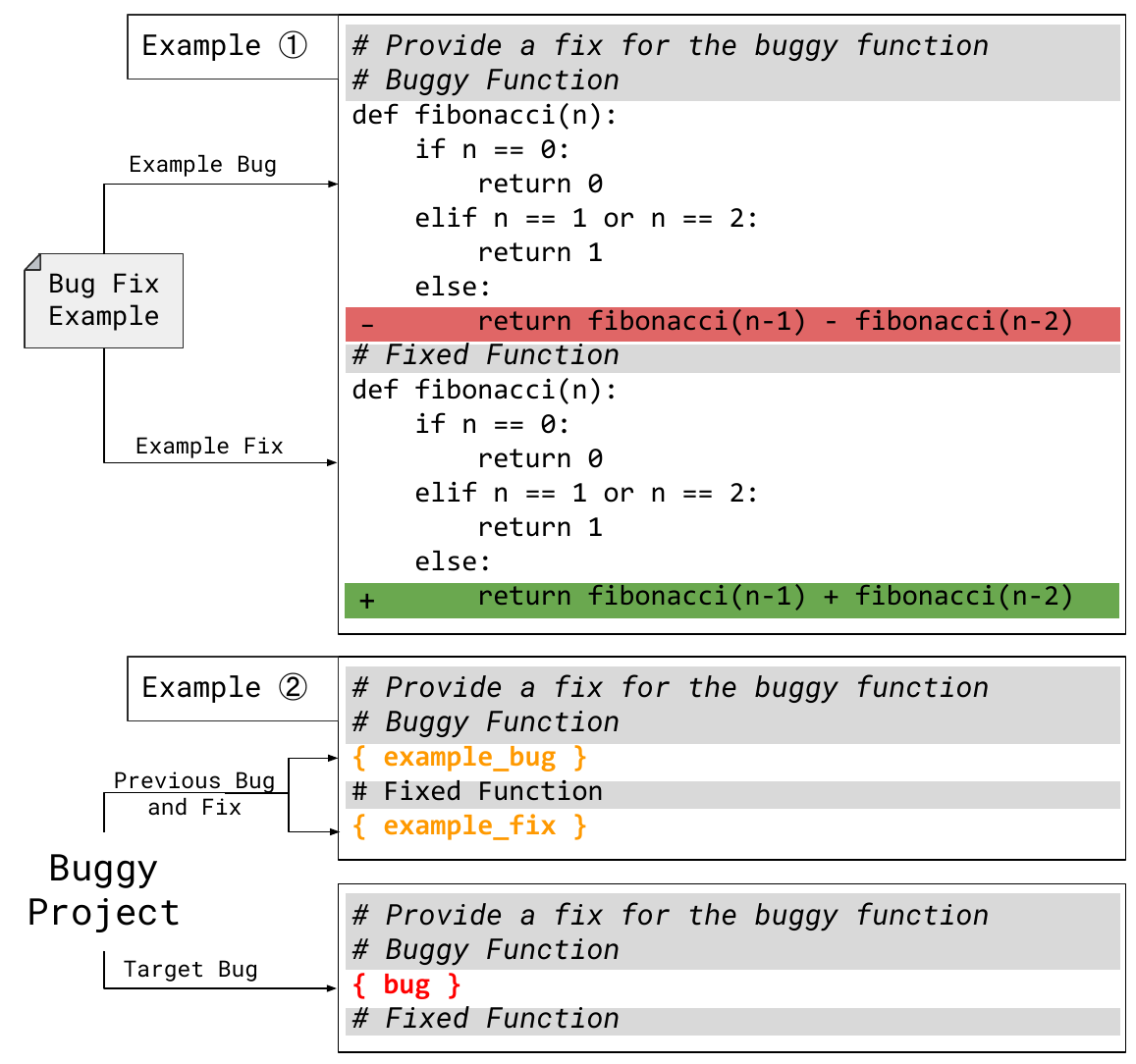}
    \centering
    \caption{\apr input for complete function generation}
    \label{fig:clm_input}
\end{figure}

\begin{figure}
    \includegraphics[width=0.7\linewidth]{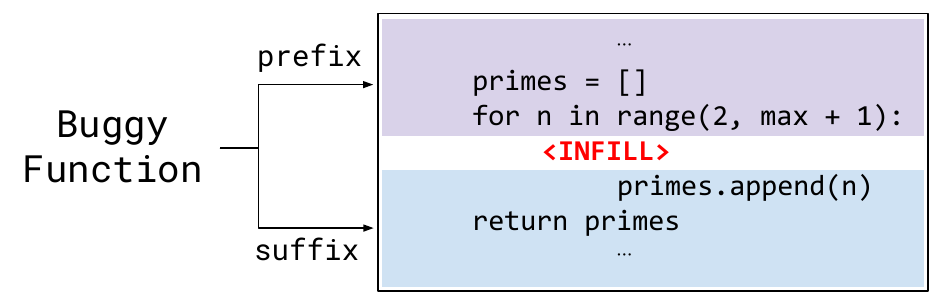}
    \centering
    \caption{\apr input for correct code infilling}
    \label{fig:mlm_input}
\end{figure}

\begin{figure}
    \includegraphics[width=\linewidth]{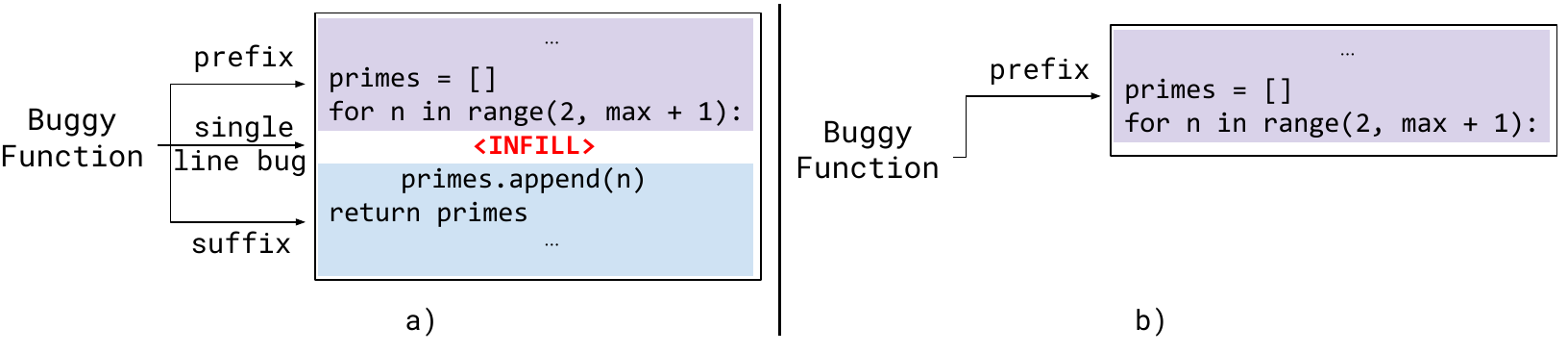}
    \caption{\apr input for single line generation}
    \label{fig:slb_input}
\end{figure}

\subsubsection{Complete function generation} \label{sec:cpf_generation}
For this setting, the initial input is the original buggy function. We aim to use a \clm to autoregressively generate the entire patched version of the buggy function. However, naively feeding the \plm{s} the buggy function will not work since each \plm is not pre-trained for \apr (i.e., they do not know that the goal is to generate a patched function). Therefore, to facilitate the direct usage of \plm{s} for \apr, we use specific prompts to enable the models to perform few-shot learning. This allows the \plm{s} to recognize the task and generate a patched function by completing the input provided. We note here that the task of complete function generation makes no assumption of 1) the location of the bug and 2) the type of bug or fix required. Therefore, the \plm needs to figure out \textit{why} the function is buggy and provide a patch to fix the bug.

Figure~\ref{fig:clm_input} shows the input which is made up of two example bug fixes (one crafted by us and one from the same project/dataset the target bug is from) in order to demonstrate the task and the expected format of the output. To start off, we follow the prior study~\cite{prenner2021codexws} and begin with a description of the task: \CodeIn{\# Provide a fix for the buggy function}. This describes in natural language the task we want the \plm to perform. This is a Python example and we use the Python comment format of \CodeIn{\#} as a prefix for this description (we use other comment prefixes depending on the language of the buggy code). We then provide an example bug and patch pair. In Figure~\ref{fig:clm_input}, this example is a function which computes the Fibonacci number. We prefix the example buggy and fixed function with \CodeIn{\# Buggy Function} and \CodeIn{\# Fixed Function} to provide additional context for the model. For our second example, we follow the same prompting style and pick a buggy and patched function pair from the same project that the bug is from. This way we can provide the model with some {examples of} the coding style used in the project. Finally, we finish the prompt by adding the bug we want to fix. %

\subsubsection{Correct code infilling} Unlike complete function generation, where the buggy location within the function is not known. For correct code infilling, the input is the prefix and suffix after removing the buggy code hunk. In order to fill in the correct code, both the prefix and suffix can provide useful information. As a result, \clm{s} are not suitable for this task since the generation process conditions only on the context to the left (prefix). Therefore, for correct code infilling, we only use \mlm{s} which perform generation by conditioning on both left (prefix) and right (suffix) code. 

Figure~\ref{fig:mlm_input} shows an example input for the infilling task. We start with the target buggy function we want to fix and remove the buggy code hunk. This gives us the prefix and suffix code which are still correct. We then place an infilling token between the prefix and suffix. This infilling token (e.g., \CodeIn{<INFILL>}) indicates to the model that this is the location where we want the new code to be generated at. The model then generates only the code to fill in the missing chunk and we obtain a patch by combining the model output with the prefix and suffix code snippets.

\subsubsection{Single line generation} In single line generation, the buggy location is provided and the bug requires only a single line change. Figure~\ref{fig:slb_input}a shows a similar setup to correct code infilling where we provide both the prefix and suffix code and use \mlm{s} to generate a replacement line. Different from correct code infilling, we can also use \clm{s} by providing only the prefix. Figure~\ref{fig:slb_input}b demonstrates the setup to use \clm{s} for this task. Since we know the bug requires only a single line change, we can stop the generation after the model has provided us with one line. We cannot apply the same strategy using \clm{s} for correct code infilling since those bugs may need multiple lines to fix and we do not know when we can stop the generation~\cite{kolak2022patch}. Additionally, when using \clm{s} for single line generation, we cannot provide the models with the suffix code due to the causal nature of the \clm{s}. We contrast this with \mlm{s} on the same task to demonstrate the effect of including the suffix context for \apr.  

\subsection{Patch Ranking and Validation} %
For all 3 repair tasks, the patch generation process is similar -- we provide the \plm{s} with the constructed input and use sampling to generate multiple patches per bug. We use nucleus sampling~\cite{holtzman2019nucleus} with {a} sampling temperature. A lower temperature means the model is likely to pick tokens with higher likelihood, resulting in samples that are more similar (temperature of 0 gives deterministic result by picking the most likely token at each generation step). A higher temperature gives more probability for the model to pick a token with a lower likelihood, leading to more unique and interesting samples. How to pick an optimal temperature value is not obvious for a problem such as \apr. For certain bugs, one may prefer a lower temperature value in order to quickly arrive at a reasonable patch. For harder bugs, a higher temperature value can be useful to generate more unique patches in an attempt to provide a fix. For our experiments, we use the default setting used in previous work~\cite{chen2021codex, fried2022incoder}. 

In addition to generating patches, we also record the entropy value of each patch. Entropy captures how \textit{natural}~\cite{hindle2012softwarenatural} the generated sample is according to the model and can be calculated as the negative log probability of each generated token. Let \({t_1, t_2, ..., t_n}\) be the list of tokens generated and \(p_{t_i}\) be the model probability of generating token \(t_i\) given the previous context and generated tokens. Entropy is defined as:

\begin{align}
\label{eq:entropy}
    \mathit{mean\_entropy}&=\minus\!\sum_{i=1}^{n} \frac{\log(p_{t_i})}{n}\\
    \mathit{sum\_entropy}&=\minus\!\sum_{i=1}^{n} \log(p_{t_i})
\end{align}

Mean entropy averages entropies of all tokens generated whereas sum entropy computes the total entropy of the sequence. For patch ranking, we prioritize patches with lower entropy first. In this way, patches that are more \emph{natural}~\cite{hindle2012softwarenatural} can be ranked higher. Previous work on leveraging \plm{s} for \apr either used mean entropy~\cite{xia2022alpharepair} or sum entropy~\cite{kolak2022patch} without thorough evaluation, and mainly focused on patch ranking. In contrast, in this work, we empirically compare both entropy computations, and have further applied them for both patch ranking and patch correctness checking~\cite{wang2022pcc}.
Finally, for each patch generated, we filter out any patches with syntactic or semantic errors and validate the rest against the test suites to identify patches which pass all the tests.

\section{Experimental Setup}

\subsection{Research Questions} 

We study the following research questions:
\begin{itemize}[leftmargin=*]
    \item \textbf{RQ1: How do different types of \plm{s} perform for different \apr settings?} We study the effectiveness of different \plm{s} on different repair datasets, across different languages and on different \apr tasks. Furthermore, we evaluate the scaling behavior of \plm{s} when increasing model size with respect to \apr ability, computation time and compilation rates to holistically evaluate each \plm. 
    \item \textbf{RQ2: How does directly applying \plm{s} for \apr compare against state-of-the-art \apr tools?} We compare the results using \plm{s} against state-of-the-art baselines. We study the unique bugs fixed by \plm{s} and highlight the advantages of directly applying \plm{s} for \apr. 
    \item \textbf{RQ3: Can \plm{s} be directly used for patch ranking and correctness checking?} We use the built-in naturalness metric of \plm{s} (entropy) to evaluate if \plm{s} considers patched functions to be \textit{more natural} than buggy functions and if entropy can directly rank the patches for patch ranking and correctness checking.
    \item \textbf{RQ4: Can we further improve the performance of \plm{s}?} We explore two directions for further improving \plm{s}' performance for \apr: 1) increasing the number of samples, and 2) combining \plm{s} with templates.  
\end{itemize}

\subsection{Implementation}

We implement the generation pipeline in Python using PyTorch~\cite{PyTorchWebPage} versions of each \plm. We use the Hugging Face~\cite{HuggingFaceWebPage} library to load the model weights and generate outputs. For \codex, we use API access provided by OpenAI to query the model~\cite{OpenAI} with \textit{code-davinci-002} engine. To use \codex for correct code infilling, we append the API request with an additional suffix parameter~\cite{CodexSuffixAPI} with the extracted suffix from the bug. For all our experiments, we directly reuse the weights of each model. Our default setting for generation uses nucleus sampling~\cite{holtzman2019nucleus} with top \(p\) = 0.95, temperature = 0.8 and 200 samples per bug. This generation setting is consistent with previous studies on \plm{s}~\cite{chen2021codex, fried2022incoder, kolak2022patch}.  Patches are generated on a 32-Core workstation with Ryzen Threadripper PRO 3975WX CPU, 256 GB RAM and NVIDIA RTX A6000 GPU, running Ubuntu 20.04.4 LTS. 

\subsection{Subject Systems}

\begin{table}
    \caption{Evaluation dataset statistics}
    \centering
    \label{tab:evaluation_datasets}
    \begin{tabular}{@{}lcccccc@{}}
        \toprule
        \textbf{Dataset} & \textbf{\#Bugs} & \textbf{\#SF} & \textbf{\#SH} & \textbf{\#SL} & \textbf{Source} & \textbf{Language}\\
        \midrule
        \dfj 1.2 & \dfjonetotal & \dfjonesingle & \dfjonehunk & \dfjoneline & real-world & Java \\
        \dfj 2.0 & \dfjtwototal & \dfjtwosingle & \dfjtwohunk & \dfjtwoline & real-world & Java \\
        \addlinespace[2pt]
        \makecell[l]{\quix-\\Java} & \quixjavatotal & \quixjavasingle & \quixjavahunk & \quixjavaline & \makecell{coding\\problems} & Java \\
        \addlinespace[2pt]
        \makecell[l]{\quix-\\Python} & \quixpythontotal & \quixpythonsingle & \quixpythonhunk& \quixpythonline & \makecell{coding\\problems} & Python \\
        \addlinespace[2pt]
        \manybugs & \manytotal & \manysingle & \manyhunk & \manyline & real-world & C \\
        \midrule
        \textbf{Total} & \totaltotal & \totalsingle & \totalhunk & \totalline \\ 
        \bottomrule
        
    \end{tabular}
\end{table}

For evaluation, we use 5 \apr benchmarks spanning across 3 programming languages. We focus on bugs where the fix is within a single function, which is also the focus of most recent \apr work~\cite{zhu2021recoder, jiang2021cure, lutellier2020coconut, li2022alphacode}. To this end, we filter these benchmarks to find bugs that fit our designed repair settings. Table~\ref{tab:evaluation_datasets} presents the details of each repair dataset. Column \textbf{Dataset} is the dataset name, \textbf{\#Bugs} is the total number of bugs, \textbf{\#SF, \#SH, \#SL} shows the number of bugs which the reference fix is within a single function, single hunk (consecutive lines) and single line. \textbf{Source} refers to where the bugs are collected from, \textbf{Language} is the programming language of the bugs. We next discuss the detailed dataset information:

\subsubsection{\dfj 1.2 and 2.0~\cite{just2014dfj}} The most widely studied \apr benchmark with a collection of bugs gathered from open-source projects in Java containing pairs of buggy and patch versions of the source project. Since \dfj has been updated to include more bugs from additional projects, we consider 2 different versions of \dfj. \dfj 1.2 contains \dfjonetotal bugs (removing the 4 depreciated bugs) from 6 open-source Java projects. \dfj 2.0 contains \dfjtwototal new bugs from 9 additional projects. Each bug in \dfj also contains developer tests exposing the bug.

\subsubsection{\quix-Python and -Java~\cite{lin2017quixbug}} A multi-lingual repair benchmark with \quixjavatotal classic programming problems. \quix benchmark is constructed from a programming challenge where programmers were asked to fix a small buggy function. \quix was originally in Python but has been translated to Java, with both versions having the same \quixjavatotal bugs. Each bug is accompanied with multiple test inputs and expected outputs.

\subsubsection{\manybugs~\cite{legoues2015manybugs}} A C repair dataset consisting of \manytotal bugs gathered from 9 open-source projects with developer written tests. Each bug is manually verified and classified into a bug type. However, we were not able to reproduce all bugs from the dataset (i.e., builds successfully and reference patches can pass all provided tests). As such we only use the \manyrepro bugs where the results were reproducible by us. 

\subsection{Compared Techniques}

We compare against the state-of-the-art \apr baselines with both \learning and traditional \apr tools. We choose 8 recent \learning \apr tools: \alpharepair~\cite{xia2022alpharepair}, \rewardrepair~\cite{ye2022rewardrepair}, \recoder~\cite{zhu2021recoder}, \deepdebug~\cite{drain2021deepdebug}, \cure~\cite{jiang2021cure}, \coconut~\cite{lutellier2020coconut}, \dlfix~\cite{li2020dlfix} and \sequencer~\cite{chen2018sequencer}. Apart from \alpharepair, these \learning \apr baselines are based on the \nmt models. \alpharepair combines a \plm (\codebert) with simple templates to generate patches under a zero-shot setting. Furthermore, we also choose 12 traditional \apr tools: \tbar~\cite{liu2019tbar}, \prapr~\cite{ghanbari2019prapr}, \avatar~\cite{liu2019avatar}, \simfix~\cite{jiang2018simfix}, \fixminer~\cite{koyuncu2020fixminder}, \capgen~\cite{wen2018capgen}, \jaid~\cite{chen2017jaid}, \sketchfix~\cite{hua2018sketchfix}, \nopol~\cite{demacro2014nopol}, \jgenprog~\cite{martinez2015automatic}, \jmutrepair~\cite{martinez2016astor}, and \jkali~\cite{martinez2016astor}.
In total, we evaluate against 20 different \apr tools. 
We compare against the baseline results on \dfj 1.2, 2.0, \quix-Python and Java on perfect fault localization - the ground-truth fix location is known to the repair tool. This is the preferred comparison setting as it eliminates the impact of differences in fault localization have on the result~\cite{lutellier2020coconut, jiang2021cure, zhu2021recoder, tufano2018empstudy}. Due to the lack of recent \apr tools that are evaluated on \manybugs, we only use it for RQ1. We follow prior work~\cite{zhu2021recoder, jiang2021cure, liu2019tbar, ghanbari2019prapr} and directly use the correct patch results from previous studies~\cite{xia2022alpharepair, ghanbari2019prapr, liu2019tbar}.

\subsection{Evaluation Metrics}

To evaluate the repair performance, we use the standard metrics of \textit{plausible patches} -- passing the all test cases, and \textit{correct patches} -- syntactically or semantically equivalent to the reference patches. To determine correct patches, we follow the standard practice in \apr research and manually inspect each plausible patch for semantic equivalency.

\section{Result}

\subsection{RQ1: Comparison of Different \plm{s}}

\begin{table}
 \caption{Complete function \apr}
        \centering
        \setlength{\tabcolsep}{2pt}
        \label{tab:cl_evaluation}
        \begin{threeparttable}
        \begin{tabular}{@{}lcccccc@{}}
        \toprule
        \textbf{Dataset} & \makecell{\gptneo \\ 125M} & \makecell{\gptneo \\ 1.3B} & \makecell{\gptneo \\2.7B} & \gptj & \makecell{GPT-\\NeoX} & \codex \\
        \midrule
        \dfj 1.2 & 6 / 8 & 7 / 16 & 10 / 24 & 14 / 31 & 18 / 36 & 63 / 102\\
        \dfj 2.0 & 2 / 17 & 4 / 18 & 6 / 20 & 11 / 33 & 15 / 36 & 49 / 93\\
        \quix-Java & 1 / 3 & 4 / 5 & 3 / 5 & 3 / 5 & 8 / 9 & 32 / 35\\
        \quix-Py & 1 / 3 & 4 / 6 & 4 / 6 & 13 / 17 & 19 / 22 & 37 / 37\\
        \manybugs & 0 / 2 & 1 / 4 & 2 / 4 & 3 / 6 & 4 / 12 & 7 / 15\\
        \bottomrule
    \end{tabular}
    \end{threeparttable}
\end{table}

\begin{table}
   \caption{Correct code infilling \apr}
        \centering
        \label{tab:infill_evaluation}
        \begin{tabular}{@{}lcccc@{}}
        \toprule
        \textbf{Dataset} & \codet & \makecell{\incoder 1.3B} & \makecell{\incoder 6.7B} & \codex \\
        \midrule
        \dfj 1.2 & 6 / 13 & 32 / 51 & 37 / 53 & 62 / 77\\
        \dfj 2.0 & 12 / 19 & 31 / 56 & 37 / 61 & 56 / 85\\
        \quix-Java & 10 / 10 & 21 / 26 & 26 / 29 & 34 / 36\\
        \quix-Py & 7 / 8 & 25 / 26 & 27 / 28 & 39 / 39\\
        \manybugs  & 2 / 5 & 8 / 12 & 9 / 13 & 12 / 15 \\
        \bottomrule
    \end{tabular}
\end{table}

\begin{table*}
 \caption{Single line \apr}
        \centering
        \label{tab:sl_evaluation}
        \begin{tabular}{@{}lcccccccccc@{}}
        \toprule
        \textbf{Dataset} & \makecell{\gptneo \\ 125M} & \makecell{\gptneo \\ 1.3B} & \makecell{\gptneo \\ 2.7B} & \gptj & \gptneox & \codet & \makecell{\incoder \\ 1.3B} & \makecell{\incoder\\ 6.7B} & \makecell{\codex \\ single-line} & \makecell{\codex \\ suffix} \\
        \midrule
        \dfj 1.2 & 5 / 10 & 12 / 20 & 13 / 21 & 16 / 26 & 21 / 31 & 5 / 12 & 21 / 36 & 26 / 38 & 32 / 37 & 39 / 47\\
       \dfj 2.0 & 8 / 17 & 10 / 26 & 16 / 28 & 12 / 26 & 19 / 36 & 9 / 15 & 15 / 32 & 21 / 37 & 26 / 38 & 31 / 45\\
       \quix-Java & 8 / 9 & 19 / 20 & 16 / 17 & 20 / 21 & 20 / 21 & 10 / 10 & 21 / 26 & 26 / 29 & 30 / 31 & 34 / 36\\
        \quix-Python & 9 / 10 & 14 / 14 & 22 / 23 & 26 / 27 & 28 / 28 & 7 / 8 & 25 / 26 & 27 / 28 & 36 / 36 & 39 / 39\\
        \manybugs & 2 / 4 & 2 / 5 & 3 / 5 & 6 / 7 & 6 / 9 & 2 / 4 & 8 / 11 & 9 / 11 & 8 / 10 & 10 / 11 \\
        \bottomrule
    \end{tabular}
\end{table*}

\subsubsection{Repair effectiveness}
We first compare \plm{s} against each other in generating plausible and correct patches. Table~\ref{tab:cl_evaluation} shows the results of 6 \clm{s} under complete function generation setting. The two integers in each cell represent the number of correct and plausible patches. We first observe that similar to previous studies in \nlp~\cite{kaplan2020scaling}, there is a scaling effect on the repair effectiveness. \textit{As we increase the size of the model, we also increase in the number of correct and plausible patches generated}. Directly looking at the group of GPT models trained on the same dataset, we see that the performance consistently increases as we use larger models across all repair datasets. However, we see that the \codex model (12B) outperforms the biggest model (\gptneox (20B)). We hypothesize that this is because \codex is designed and finetuned for code generation; on the other hand, while the training dataset of \gptneox is partially made up of code, it is designed for general purpose text generation. 

Tables~\ref{tab:infill_evaluation} and~\ref{tab:sl_evaluation} show the results on the correct code infilling and single line generation repair tasks. Similar to the previous result, we again see the scaling effect of increased performance as model size increases. Compared to complete function generation, we observe that each model using correct code infilling and single line generation is able to produce a higher ratio of correct fixes to the total number of bugs. Furthermore, we also observe that the ratio of correct patches to plausible patches is higher in the latter 2 settings as well. This signals that patches produced using code infilling and single line generation is more likely to be the correct fix. The improved performance is because for complete function generation the model needs to understand the prompt given (\ref{sec:cpf_generation}), localize the bug and provide the correct fix. On the other hand, when we provide the model with the buggy location information in correct code infilling and single line generation, it only needs to fill in or complete the partial code, leading to more correct patches. This comparison is more direct when evaluating the \codex model, the only model that can perform both code infilling and function generation. We see that when performing correct code infilling, \codex is able to fix 40\% (62/154) of the total bugs whereas when asked to generate the entire function, it drops to 28\% (63/225).

For single line generation results in Table~\ref{tab:sl_evaluation}, we included both generative and \mlm{s}. However, for \clm{s} we are not able to provide it with suffix code snippets since their generation is dependent only on the previous context. We compare this with \mlm{s}, which can perform infilling conditioned on both the context before and after. We observe that \mlm{s} perform better than their generative counterparts. Additionally, since we are able to use both the generative and infilling versions of \codex, we can directly compare the repair ability of the model when given only the prefix versus both prefix and suffix context. We see that when using the suffix information from the original buggy function, the \codex model is able to improve the number of correct and plausible fixes across all repair datasets. This shows that for repair, \emph{successfully utilizing the code after the buggy lines is important for fixing bugs}.

\begin{table}
\caption{Patch generation speed (\#patch/min)}
        \centering
        \label{tab:speed}
        \begin{tabular}{@{}l rrr rrr@{}}
        \toprule
        \textbf{Models} & \multicolumn{3}{c}{\textbf{\dfj 1.2}} & \multicolumn{3}{c}{\textbf{\quix-Python}} \\
        \cmidrule(lr){2-4}\cmidrule(l){5-7}
        & \textbf{CF} & \textbf{CI} & \textbf{SL} & \textbf{CF} & \textbf{CI} & \textbf{SL} \\
        \midrule
        \gptneo 125M & 139 & -  & 1080 & 369 & -  & 1061\\
        \gptneo 1.3B & 31 & -  & 543 & 127 & -  & 814\\
        \gptneo 2.7B & 27 & -  & 489 & 85 & -  & 625\\
        \gptj & 15 & -  & 227 & 39 & -  & 354\\
        \gptneox & 2 & -  & 47 & 6 & -  & 73\\
        \codet & -  & 969 & -  & -  & 1991 & - \\
        \incoder 1.3B & -  & 535 & -  & -  & 1083 & - \\
        \incoder 6.7B & -  & 288 & -  & -  & 419 & - \\
        \bottomrule
        \end{tabular}
\end{table}

\subsubsection{Speed} Next we look at the speed of patch generation using \plm{s}. We already saw from the previous result analysis that as we increase the size of the model, we obtain an increase in repair performance. However, such performance increase does not come for free as larger models require longer time for inferencing. Table~\ref{tab:speed} shows the samples generated per minute for different \plm{s} on \dfj 1.2 and \quix-Python with the 3 repair generation settings (Columns CF, CI, SL refer to complete function, correct infilling and single line generation, respectively). We only include models that we run locally on the same hardware (i.e., excluding \codex since it is only accessible through API access). We first observe that as we increase model size, the patch generation speed drastically slows down (71x slower on \gptneox than \gptneo 125M on complete function generation). This demonstrates the trade-off between repair effectiveness and time cost when using large models. Additionally, we see that compared to single line generation and correct code infilling, complete function generation takes significantly more time, since generating an entire function is much more time consuming than generating a single line or hunk. This shows that when the buggy location is known, we can use infilling and single line generation settings to achieve faster repair.

\begin{figure}
    \centering
    \includegraphics[width=0.9\linewidth]{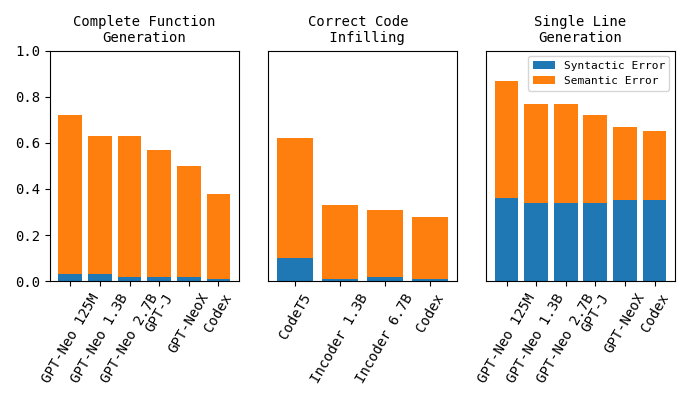}
    \caption{Syntactic and semantic error rates on \dfj 1.2}
    \label{fig:compile_comp}
\end{figure}

\subsubsection{Compilation rate} We evaluate the compilation rate of the patches generated by each \plm{}. Figure~\ref{fig:compile_comp} shows the syntactic and semantic error rates of all studied \plm{s} using the three repair settings on \dfj 1.2. We first observe that the overall error rate (syntactic + semantic) of the generated patches goes down as we increase the size of the model. This reaffirms the previously discussed scaling effect of \plm{s} and show that the patches generated by larger models contain less errors. Next we see that all \clm{s} using single line generation produced a high number of syntactic errors. Recall that single line generation when using \clm{s} only provides the prefix in the buggy function. As a result, the generated line can easily introduce some syntax errors (e.g., adding an if statement with an opening bracket) since the model does not know what the suffix code context is. On the other hand, the amount of syntax errors produced in the two other settings are much lower. For complete function generation, \plm{s} can effectively retain the syntax of the language during training and generate syntactically correct functions. \emph{For correct code infilling, not only do we get low syntactic errors but also achieve the lowest semantic errors.} Having both the prefix and suffix provides the model with sufficient context which leads to higher compilable patch rate. 

\subsection{RQ2: Comparison against State-of-the-art \apr tools}

\begin{table}
\caption{\dfj 1.2 baseline comparison}

        \centering
        \label{tab:comparison_d4j12}
        \begin{tabular}{@{}lcccc@{}}
        \toprule
        \textbf{Tools / Models} & \textbf{\makecell{Single func. \\ (\dfjonesingle bugs)}} & \textbf{\makecell{Patch \\ func.}} & \textbf{\makecell{Correct \\ hunk}} & \textbf{\makecell{Single \\ line}} \\  
        \midrule
        \alpharepair & 67 \\
        \rewardrepair & 48\\
        \recoder & 61 \\
        \tbar & 54 \\
        \cure & 52\\
        \midrule
        \gptneo 125M & 9 & 6 & -  & 5\\
        \gptneo 1.3B & 18 & 7 & -  & 12\\
        \gptneo 2.7B & 20 & 10 & -  & 13\\
        \gptj & 28 & 14 & -  & 16\\
        \gptneox & 34 & 18 & -  & 21\\
        \codet & 6 & -  & 6 & - \\
        \incoder 1.3B & 32 & -  & 32 & - \\
        \incoder 6.7B & 37 & -  & 37 & - \\
        \codex & 99 & 63 & 62 & 32\\
        \midrule
        Total  & 109 & 69 & 74 & 40\\
        \bottomrule
        \end{tabular}
\end{table}

\subsubsection{\dfj 1.2 results}
We first compare the results of directly using \plm{s} for repair against both traditional and \learning \apr tools on \dfj 1.2. Table~\ref{tab:comparison_d4j12} shows the number of correct bug fixes of the top baseline tools and also the \plm{s} in our evaluation. The last 3 columns present the number of correct patches generated when using each of the three \apr settings. We then combine all patches generated for each of the models together (Column 2) to demonstrate the total number of fixes that can be obtained for the 255 single function bugs in \dfj 1.2. Note that this is still a fair comparison -- prior \apr techniques typically use a timeout of 5h for each bug~\cite{zhu2021recoder, jiang2021cure, xia2022alpharepair}, while generating 200 patches for each of the 3 settings (i.e., at most 600 patches in total) costs no more than 2.5 hours for each model.  

We observe that {some} of the models are able to achieve comparable performances compared to some of the recent state-of-the-art \apr tools. Additionally, this result is obtained while generating only {up to 600} samples per bug whereas prior approaches, especially \learning tools, can generate up to 5000 patches per bug~\cite{lutellier2020coconut, jiang2021cure, xia2022alpharepair}. While the most effective model (\codex) can already outperform all existing techniques (e.g., fixing 99 single-function bugs), by combining the patches generated by all models (Total), we can achieve 109 correct fixes on single function bugs! The surprising results show that \textit{by directly applying \plm{s} for \apr without any specific change/finetuning, we can already achieve the highest number of correct fixes compared to existing baselines.}

\begin{figure}
    \centering
    \includegraphics[width=0.9\linewidth]{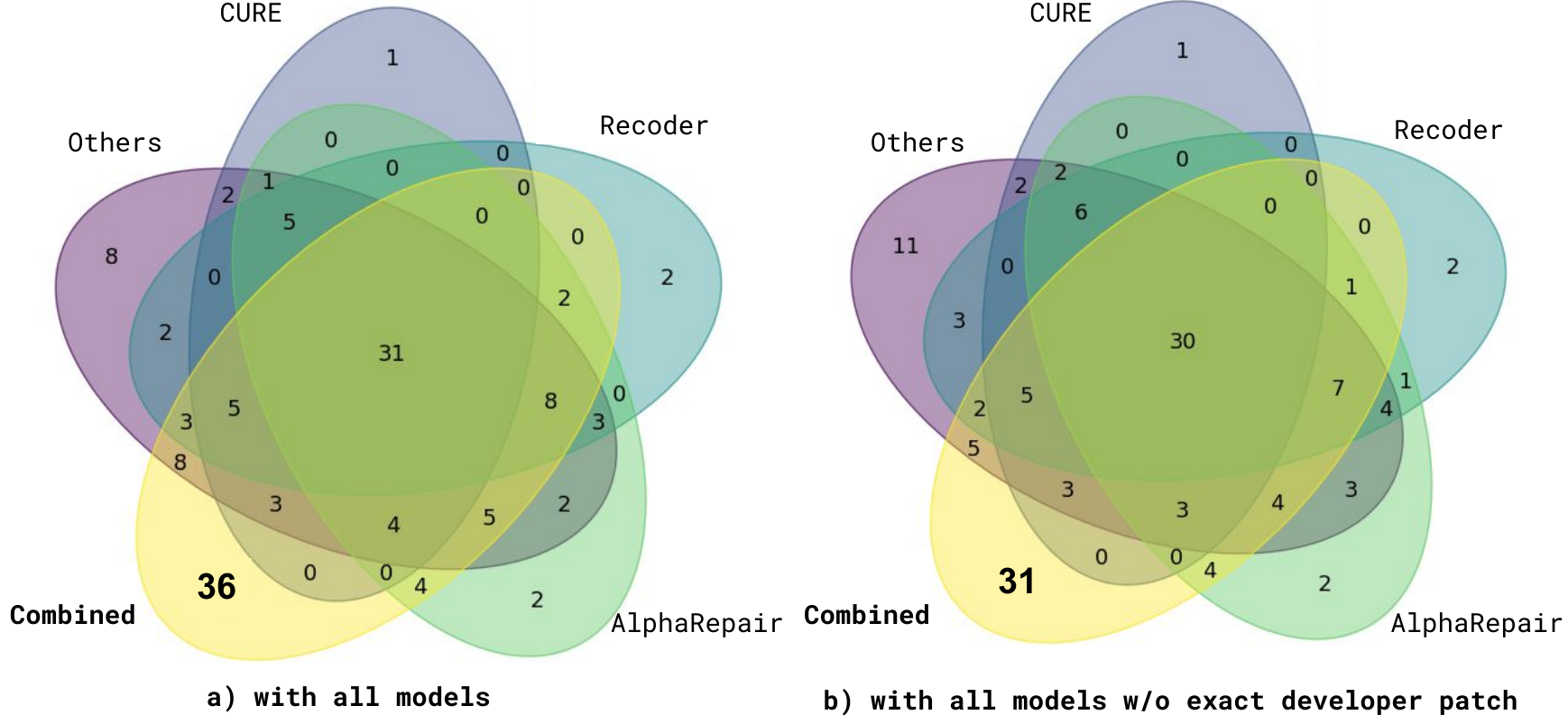}
    \caption{Bug fix Venn diagram on \dfj 1.2}
    \label{fig:venn_dfj12}
\end{figure}

\begin{figure}
    \includegraphics[width=0.8\linewidth]{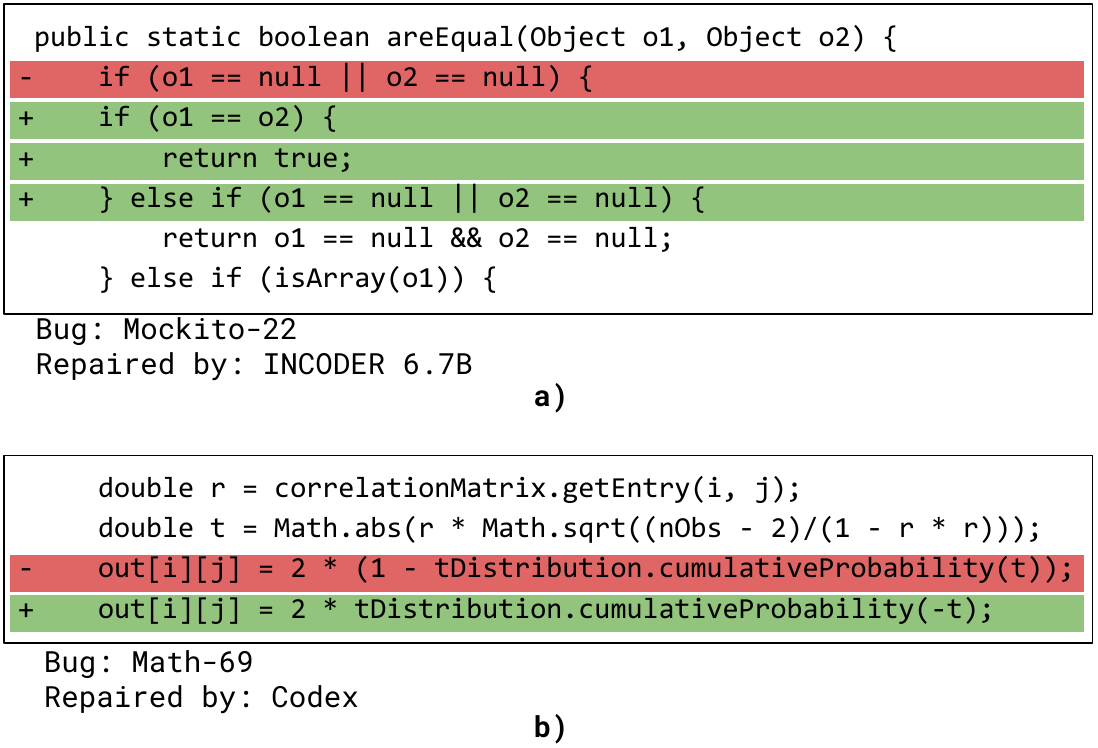}
    \centering
    \caption{Unique bug fixes generated by \plm{s}}
    \label{fig:example_fix_dfj12}
\end{figure}

Figure~\ref{fig:venn_dfj12} presents the Venn diagram of unique fixes that can be generated using \plm{s} compared to the 3 best performing baselines on all the single function bugs in \dfj 1.2. We also combine all fixes from other baselines together into the ``Others'' category in the Venn diagram. We observe that by combining all the models together, we can  generate a significant amount of unique bug fixes (36) that no other tools have fixed so far. Due to the potential data leakage issue (discussed in detail in Section~\ref{sec:threats}), we further investigate whether \plm{s} can generate correct patches that are not exactly the same as developer patches. Figure~\ref{fig:venn_dfj12}b shows the unique bug fixes on \dfj 1.2 compared to the baselines when we remove all fixes which are exactly the same as the developer patch. We observe that combining all \plm{s} together would still achieve the highest number of bug fixes (93) with 31 unique bug fixes. 

To demonstrate the ability of these \plm{s}, we show some unique fixes produced by them. Figure~\ref{fig:example_fix_dfj12}a is a correct patch produced by the \incoder 6.7B model under correct code infilling task. We see here that the function is called \CodeIn{areEqual} and the bug is caused by missing a specific case of comparing if the two inputs have the same reference. Using both the prefix (name of the function) and suffix (other comparison statements with return values), the model figures out the correct code to be insert here (first checking if the references are the same before proceeding). Such code is commonly found in open-source projects which uses similar comparison functions where the \plm{s} can learn from. In fact, we found several similar comparison functions (checking if the objects have the same reference)~\cite{pilefix1, pilefix2, pilefix3, pilefix4} in different projects as a part of The Pile dataset~\cite{gao2020pile} that some of the \plm{s} were trained on. Furthermore, unlike traditional \apr tools which often work on a single line, \plm{s} can generate multiple lines of code in order to provide the correct fixes. %

Figure~\ref{fig:example_fix_dfj12}b shows a patch of the Math-69 bug generated by \codex. The function here calculates a matrix of p-values of a 2-sided, 2-sample t-test. The bug is caused by precision error when the function call is extremely close to 1. Here the model generates an alternative way of calculating the p-value which is much more stable than before. This is a hard bug to fix since the change is quite subtle however it does not fit any of the common templates used in traditional \apr. To generate the correct fix, the model needs to understand the goal of the function (p-value calculation) and use statistical formulas. Both of which is in \codex as it is trained not only on code but also on general text, which contains many descriptions and examples of t-test p-value calculations. This unique fix shows the benefit of using \plm{s} for program repair where domain knowledge of the project can be utilized as well. 

\subsubsection{Additional results} 

\begin{table}
\caption{Additional baseline comparison}
        \centering
        \label{tab:comparison_add}
        \begin{tabular}{@{}lccc@{}}
        \toprule
        \textbf{Tools / Models} & \textbf{\makecell{\dfj 2.0 \\ (78 bugs)}} & \textbf{\makecell{\quix-Java \\ (40 bugs)}} & \textbf{\makecell{\quix-Python \\ (40 bugs)}}\\
        \midrule
        \alpharepair & 35 & 28 & 27\\
        \rewardrepair & 25 & 20 & -\\
        \deepdebug & - & - & 21\\
        \recoder & 11 & 17 & -\\
        \cure & - & 21 & -\\
        \tbar & 8 & - & -\\
        \coconut & - & 13 & 19\\
        \midrule
       \gptneo 125M & 10 & 8 & 9\\
        \gptneo 1.3B & 11 & 20 & 17\\
        \gptneo 2.7B & 19 & 18 & 24\\
        \gptj & 16 & 22 & 29\\
        \gptneox & 24 & 21 & 31\\
        \codet & 9 & 10 & 7\\
        \incoder 1.3B & 15 & 21 & 25\\
        \incoder 6.7B & 21 & 26 & 27\\
        \codex & 45 & 38 & 40\\
        \midrule
        Total  & 52 & 38 & 40\\
        \bottomrule
        \end{tabular}
\end{table}

In addition to comparing against state-of-the-art baselines on \dfj 1.2, we also compare the performance of \plm{s} on other datasets widely used to evaluate previous \apr tools. Table~\ref{tab:comparison_add} shows the results on \dfj 2.0, \quix-Java and -Python where we also combine the correct bug fixes of the 3 generation strategies together. Similar to the \dfj 1.2 results, we observe that many models can achieve similar (or even better) performance with carefully designed \apr tools. More surprisingly, \emph{all 9 studied \plm{s} can outperform \tbar, state-of-the-art template-based \apr tool, and are competitive compared with the recent \recoder technique on the \dfj 2.0 dataset.} Furthermore, unlike many baselines which can only be used on a single language (specifically designed for a particular language or requiring additional finetuning on another language), the \plm{s} can be directly applied for multi-lingual repair. %

\subsection{RQ3: Patch Ranking and Correctness Checking Analysis}

\begin{table}
\caption{Mean entropy of generated patches}
        \centering
        \label{tab:entropy}
        \begin{tabular}{@{}c@{\quad }lrrrrrr@{}}
        \toprule
        & \textbf{Models} & \multicolumn{3}{c}{\textbf{\dfj 1.2}} & \multicolumn{3}{c}{\textbf{\quix-Python}} \\
        \cmidrule(lr){3-5}
        \cmidrule(lr){6-8}
        &  & \textbf{C} & \textbf{P} & \textbf{NP} & \textbf{C} & \textbf{P} & \textbf{NP} \\
        \midrule
        \multirow{6}{*}{\STAB{\rotatebox[origin=c]{90}{\textsl{ Function Gen.}}}} 
        & \gptneo 125M & 0.08 & 0.13 & 0.23 & 0.10 & 0.10 & 0.20 \\
        & \gptneo 1.3B & 0.12 & 0.12 & 0.19 & 0.06 & 0.05 & 0.09 \\
        & \gptneo 2.7B & 0.09 & 0.13 & 0.17 & 0.05 & 0.06 & 0.08 \\
        & \gptj& 0.07 & 0.10 & 0.12 & 0.04 & 0.05 & 0.08 \\
        & \gptneox & 0.08 & 0.11 & 0.13 & 0.05 & 0.07 & 0.10 \\
        & \codex & 0.04 & 0.05 & 0.08 & 0.11 & 0.13 & 0.16 \\
        \midrule
        \multirow{4}{*}{\STAB{\rotatebox[origin=c]{90}{\textsl{ Infilling}}}} 
        & \codet & 0.50 & 0.51 & 0.54 & 0.51 & 0.50 & 0.59 \\ 
        & \incoder 1.3B & 0.49 & 0.58 & 0.65 & 0.54 & 0.56 & 0.65 \\
        & \incoder 6.7B & 0.45 & 0.50 & 0.61 & 0.61 & 0.60 & 0.65 \\
        & \codex & 0.43 & 0.43 & 0.50 & 0.32 & 0.33 & 0.42 \\
        \midrule
        \multirow{6}{*}{\STAB{\rotatebox[origin=c]{90}{\textsl{ Line Gen.}}}} 
        & \gptneo 125M & 0.38 & 0.42 & 0.58 & 0.41 & 0.45 & 0.61 \\
        & \gptneo 1.3B & 0.32 & 0.38 & 0.58 & 0.25& 0.27 & 0.47 \\
        & \gptneo 2.7B & 0.28 & 0.32 & 0.55 & 0.21 & 0.26 & 0.40 \\
        & \gptj & 0.29 & 0.33 & 0.54 & 0.20 & 0.22 & 0.38 \\
        & \gptneox & 0.39 & 0.42 & 0.71 & 0.26 & 0.28 & 0.55 \\
        & \codex & 0.19 & 0.28 & 0.57 & 0.18 & 0.23 & 0.60 \\
        \bottomrule
        \end{tabular}
        
\end{table}

\subsubsection{Entropy}

As we are using \plm{s} for patch generation, this {allows us to compute the entropy of each patch}. Entropy calculates how natural the generated sample is (Equation~\ref{eq:entropy}). Table~\ref{tab:entropy} shows the mean entropy values for correct (\textbf{C}), plausible (\textbf{P}) and non-plausible patches (\textbf{NP}). Each row shows the results of a \plm on a repair scenario containing bugs for which the \plm can produce a correct patch. We observe that average entropy value of correct and plausible patches for all models are less than non-plausible patches. Although not shown in the table, we observe the same finding when comparing patches using sum entropy. In other words, \emph{the studied \plm{s} consider correct patches which correctly fix the underlying bugs to be more \textit{nature} than other patches}. Additionally, while the entropy difference between correct and plausible patches are not as drastic as compared to non-plausible patches, we also find that correct patches are in general less entropic than plausible ones. Recent work~\cite{wang2022pcc} has shown that existing solutions for \emph{patch-correctness checking} (i.e., identifying correct patches from plausible patches) can suffer from dataset overfitting and performance drops when applied on more complicated patches. We demonstrate for the first time that \emph{entropy computation via \plm{s} can help distinguish correct patches from plausible patches}, indicating a promising future of directly leveraging the \plm entropy metric for patch-correctness checking.

\subsubsection{Patch ranking}

\begin{figure*}
    \centering
    \includegraphics[width=0.95\linewidth]{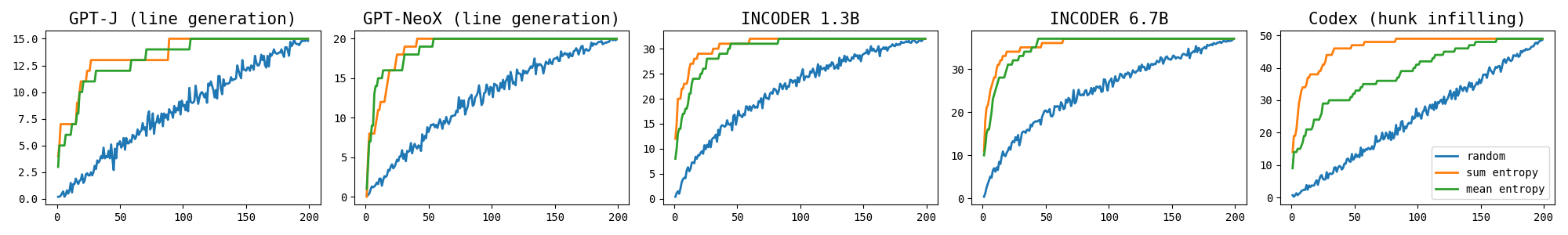}
    \caption{Number of bugs fixes when using different patch ranking strategies on \dfj 1.2}
    \label{fig:patch_ranking}
\end{figure*}

Using the entropy values of each generated patch, we perform ranking to validate patches with higher rank (lower entropy) first. We pick 5 \plm{s} with the highest number of correct patches to perform this analysis. Figure~\ref{fig:patch_ranking} shows the number of bugs fixed for the \dfj 1.2 dataset using different patch ranking strategies as we increase the number of patches to validate. We see that compared to randomly picking patches to validate (blue line), when using entropy rankings (orange and green line), we can validate the correct patches faster. This shows that entropy can be an effective measure used to rank the potential patches to prioritize lower entropy patches for validation under tighter time constraints. Furthermore, we observe that \emph{sum entropy performs slightly better compared to mean entropy}. We hypothesize that this is because sum entropy calculates the entire sequence entropy regardless of the length of the generated sequence. As such, shorter sequences tend to have lower sum entropy compared to longer sequences; interestingly, this is consistent with traditional \apr or patch correctness checking techniques~\cite{asad2019impact, xin2017leveraging, le2017s3}, which favor simple patches over complicated ones.

\subsection{RQ4: Improvements on Direct \plm \apr}

\begin{table}
\caption{Baseline comparison with \plm improvements}
        \centering
        \label{tab:template}
        \begin{tabular}{@{}lccc@{}}
        \toprule
        \textbf{Tools / Models} & \textbf{\makecell{\dfj \\1.2 All}}  & \textbf{\makecell{\dfj 2.0 \\ Single Line}}  & \textbf{\makecell{\quix-\\Python}}\\
        \midrule
        \alpharepair & 74 & 35 & 27\\
        \rewardrepair & 45 & 25 & -\\
        \deepdebug & - & - & 21 \\
        \recoder & 65 & 11 & -\\
        \tbar & 68 & 8 & -\\
        \midrule
        \makecell[l]{\incoder (200)} & 37 & 21 & 27\\ 
        \makecell[l]{\incoder (2000)} & 64 & 25 & 32\\
        \addlinespace[1pt]
        \makecell[l]{\incoder w/ template (2000)} & \textbf{78} & \textbf{39} & \textbf{37}\\
        \bottomrule
        \end{tabular}
\end{table}

In previous RQs, we showed that by directly applying \plm{s} for \apr we can already achieve comparable performance with previous \apr tools. We further explore the possibilities to boost the ability of \plm{s} for \apr. For this experimental setup, we choose the best performing model (apart from \codex, which already outperforms existing \apr techniques without any further extension) -- \incoder 6.7B and run the model longer (2000 samples per bug) combined with repair templates. We evaluate on all bugs in \dfj 1.2 by adjusting our repair settings to generate patches for every location which is changed by the reference patch instead of only on a single change location. This setup is similar to previous \learning repair tools~\cite{xia2022alpharepair, zhu2021recoder} and allows us to compare on the full \dfj 1.2 dataset. Furthermore, following prior work~\cite{xia2022alpharepair}, we include evaluation on \dfj 2.0 single line bugs and \quix-Python.

Table~\ref{tab:template} shows the baseline tools along with our model setups. \textbf{\incoder (200)} is our default setup from previous evaluation that generates 200 samples per bug. \textbf{\incoder (2000)} shows the results when we increase the number of samples to 2000. \textbf{\incoder w/ template (2000)} contains the results when combining repair template with the \incoder model. Using similar design as the \alpharepair baseline, we apply different repair templates by using the original buggy line. Such templates include: keeping parts of the prefix or suffix, replacing method calls or parameters, and changing/adding new boolean conditions or operators to the buggy line. These repair templates make use of the original buggy line and provide important starting code for the model. 

We observe that if we \textit{apply the model longer and generate more samples, we can drastically improve the number of correct bugs fixed} in all three datasets and achieve very close result to that obtained by the best baseline. Moreover, we can obtain further improvements by using simple repair templates and achieve the highest number of correctly fixed bugs on all datasets, e.g., fixing 78 bugs on \dfj 1.2 with 15 unique bug fixes that no other baseline tools have fixed before. This finding shows that not only can \plm{s} be effective when directly used for program repair, \textit{we can combine them with more domain specific techniques such as simple repair templates to further improve their performance.}

\section{Threats to Validity}
\label{sec:threats}
\mypara{Internal.} One internal threat to validity comes from our manual validation of plausible patches to determine semantically correct patches. To address this, we carefully performed the analysis and released the correct patches and code used to perform the experiments for public evaluation~\cite{correctPatchandDataset}.

Another internal threat comes from the potential data leakage of real developer patched functions being part of the original training data. To address this, we examine the patches \plm{s} generated for \dfj 1.2 since this is the most widely studied dataset for \apr and we mainly compared with state-of-the-art \apr tools on this dataset. We first check if the bugs fixed by each \plm contain correct patches different than the reference developer patches. Out of the 354 individual bug fixes by all models on \dfj 1.2, 234 fixes (66\%) contain a patch that is different than the developer patch. We also found that due to the simplicity of single line patches, majority of the correct patches generated for single line bugs are the same as the developer patch. If we exclude single line bugs, the percentage increases to 77\% (196/255). Out of the 109 bugs that can be fixed by combining all correct patches generated by all \plm{s} together (Total row in Table~\ref{tab:comparison_d4j12}), 93 bugs (85\%) are fixed by at least one correct patch that is different than the original developer patch, e.g., as shown in Figure~\ref{fig:venn_dfj12}b, removing \plm{} fixes that are exactly the same as the developer patches can still fix 31 bugs that prior tools cannot fix. %

Since we only have access to the training data used in \codet, \gptneo, \gptj and \gptneox models, we further check if the fixed function is within the training datasets when the correct patch is equivalent to the developer fix for these models. We found that while 38\% (48/128) of bugs fixes contain only the same fix as the developer patch, only 15\% (20/128) of those patches are also found in the original training data, showing that the majority of correct bug fixes provided by these \plm{s} are not simply from memorizing the training data. Moreover, our RQ4 shows that improvements can be further made by combining repair templates with \plm{s}, which is orthogonal to the data leakage issue. Additionally, We observe that \plm{s} are able to achieve the state-of-the-art results on \quix dataset which is not part of the training data as it has low number of stars on GitHub and contains synthetic bugs and patches that are not part of any larger real-world projects. Further reducing the data leakage issue would require retraining the \plm{s}, which could be extremely costly.

\mypara{External.} We evaluate \plm{s} on 5 repair datasets across 3 programming languages, making our evaluation one of the most comprehensive studies in \apr. However, our findings may still not generalize to other datasets or languages. 
\section{Conclusion}

We present an extensive evaluation on \plm{s} for direct program repair. We use 9 state-of-the-art \plm{s} with 5 different repair datasets and design different practical repair settings to compare and contrast the repair effectiveness of different \plm{s}. In our evaluation, we shed light on the scaling effect that increasing model size has on various important factors in \apr such as the number of bugs fixed, the speed of patch generation, and the compilation rate. Also, we compare the performance of \plm{s} against state-of-the-art \apr tools and highlight the unique fixes and advantages of using \plm{s} for \apr. Furthermore, we evaluated the ability for \plm{s} to perform patch ranking in order to prioritize correct patches for faster validation. Lastly, we demonstrate the possibilities (i.e., increasing the sample size and combining \plm{s} with repair templates) to further boost the performance of \plm{s} for \apr. The results from our study demonstrate promising future of adopting \plm{s} for \apr.

\bibliographystyle{IEEEtran}
\bibliography{reference}

\begin{thebibliography}{10}
\providecommand{\url}[1]{#1}
\csname url@samestyle\endcsname
\providecommand{\newblock}{\relax}
\providecommand{\bibinfo}[2]{#2}
\providecommand{\BIBentrySTDinterwordspacing}{\spaceskip=0pt\relax}
\providecommand{\BIBentryALTinterwordstretchfactor}{4}
\providecommand{\BIBentryALTinterwordspacing}{\spaceskip=\fontdimen2\font plus
\BIBentryALTinterwordstretchfactor\fontdimen3\font minus
  \fontdimen4\font\relax}
\providecommand{\BIBforeignlanguage}[2]{{%
\expandafter\ifx\csname l@#1\endcsname\relax
\typeout{** WARNING: IEEEtran.bst: No hyphenation pattern has been}%
\typeout{** loaded for the language `#1'. Using the pattern for}%
\typeout{** the default language instead.}%
\else
\language=\csname l@#1\endcsname
\fi
#2}}
\providecommand{\BIBdecl}{\relax}
\BIBdecl

\bibitem{healthcare}
K.~Luzniak, ``Software for the healthcare industry: what is it and why it’s
  worth using?'' \emph{neoteric}, 2022,
  \url{https://neoteric.eu/blog/software-for-the-healthcare-industry-what-is-it-and-why-its-worth-using}.

\bibitem{transportation}
N.~Mayersohn, ``Data driving new approaches to transportation,'' \emph{The New
  York Times}, 2022,
  \url{https://www.nytimes.com/2020/02/05/technology/data-micromobility-electric-scooters-mds.html}.

\bibitem{bug_safety}
E.~Richards, ``Software's dangerous aspect,'' \emph{The Washington Post}, 1990,
  \url{https://www.washingtonpost.com/archive/politics/1990/12/09/softwares-dangerous-aspect/9b2e9243-8deb-4ac7-9e8f-968de0806e5e/}.

\bibitem{bug_loss}
S.~Matteson, ``Report: Software failure caused \$1.7 trillion in financial
  losses in 2017,'' \emph{TechRepublic}, 2018,
  \url{https://www.techrepublic.com/article/report-software-failure-caused-1-7-trillion-in-financial-losses-in-2017/}.

\bibitem{debuggingtime}
D.~H. O'Dell, ``The debugging mindset,'' \emph{acmqueue}, 2017,
  \url{https://queue.acm.org/detail.cfm?id=3068754/}.

\bibitem{gazzola2019aprsurvey}
L.~Gazzola, D.~Micucci, and L.~Mariani, ``Automatic software repair: A
  survey,'' \emph{IEEE Transactions on Software Engineering}, vol.~45, no.~1,
  pp. 34--67, 2019.

\bibitem{legoues2012genprog}
C.~Le~Goues, T.~Nguyen, S.~Forrest, and W.~Weimer, ``Genprog: A generic method
  for automatic software repair,'' \emph{IEEE Transactions on Software
  Engineering}, vol.~38, no.~1, pp. 54--72, 2012.

\bibitem{le2016hdrepair}
X.~B.~D. Le, D.~Lo, and C.~Le~Goues, ``History driven program repair,'' in
  \emph{2016 IEEE 23rd International Conference on Software Analysis,
  Evolution, and Reengineering (SANER)}, vol.~1, 2016, pp. 213--224.

\bibitem{wen2018capgen}
M.~Wen, J.~Chen, R.~Wu, D.~Hao, and S.-C. Cheung, ``Context-aware patch
  generation for better automated program repair,'' in \emph{Proceedings of the
  40th International Conference on Software Engineering}, ser. ICSE '18, 2018,
  p. 1–11.

\bibitem{mechtaev2016angelix}
S.~Mechtaev, J.~Yi, and A.~Roychoudhury, ``Angelix: Scalable multiline program
  patch synthesis via symbolic analysis,'' in \emph{Proceedings of the 38th
  International Conference on Software Engineering}, ser. ICSE '16, 2016, p.
  691–701.

\bibitem{le2017s3}
X.-B.~D. Le, D.-H. Chu, D.~Lo, C.~Le~Goues, and W.~Visser, ``S3: syntax-and
  semantic-guided repair synthesis via programming by examples,'' in
  \emph{Proceedings of the 2017 11th Joint Meeting on Foundations of Software
  Engineering}, 2017, pp. 593--604.

\bibitem{demacro2014nopol}
F.~DeMarco, J.~Xuan, D.~Le~Berre, and M.~Monperrus, ``Automatic repair of buggy
  if conditions and missing preconditions with smt,'' in \emph{Proceedings of
  the 6th International Workshop on Constraints in Software Testing,
  Verification, and Analysis}, ser. CSTVA 2014, 2014, p. 30–39.

\bibitem{hua2018sketchfix}
J.~Hua, M.~Zhang, K.~Wang, and S.~Khurshid, ``Sketchfix: A tool for automated
  program repair approach using lazy candidate generation,'' in
  \emph{Proceedings of the 2018 26th ACM Joint Meeting on European Software
  Engineering Conference and Symposium on the Foundations of Software
  Engineering}, ser. ESEC/FSE 2018.\hskip 1em plus 0.5em minus 0.4em\relax
  {ACM}, 2018, p. 888–891.

\bibitem{martinez2016astor}
M.~Martinez and M.~Monperrus, ``Astor: A program repair library for java
  (demo),'' in \emph{Proceedings of the 25th International Symposium on
  Software Testing and Analysis}, ser. ISSTA 2016.\hskip 1em plus 0.5em minus
  0.4em\relax New York, NY, USA: Association for Computing Machinery, 2016, p.
  441–444.

\bibitem{koyuncu2020fixminder}
A.~Koyuncu, K.~Liu, T.~F. Bissyand{\'{e}}, D.~Kim, J.~Klein, M.~Monperrus, and
  Y.~L. Traon, ``Fixminer: Mining relevant fix patterns for automated program
  repair,'' \emph{Empir. Softw. Eng.}, vol.~25, no.~3, pp. 1980--2024, 2020.

\bibitem{liu2019avatar}
K.~Liu, A.~Koyuncu, D.~Kim, and T.~F.~Bissyand{\'e}, ``{AVATAR:} fixing
  semantic bugs with fix patterns of static analysis violations,'' in
  \emph{Proceedings of the 26th IEEE International Conference on Software
  Analysis, Evolution, and Reengineering}.\hskip 1em plus 0.5em minus
  0.4em\relax IEEE, 2019, pp. 456--467.

\bibitem{liu2019tbar}
K.~Liu, A.~Koyuncu, D.~Kim, and T.~F. Bissyand\'{e}, ``Tbar: Revisiting
  template-based automated program repair,'' in \emph{Proceedings of the 28th
  ACM SIGSOFT International Symposium on Software Testing and Analysis}, ser.
  ISSTA 2019.\hskip 1em plus 0.5em minus 0.4em\relax New York, NY, USA: ACM,
  2019, p. 31–42.

\bibitem{ghanbari2019prapr}
A.~Ghanbari, S.~Benton, and L.~Zhang, ``Practical program repair via bytecode
  mutation,'' in \emph{Proceedings of the 28th ACM SIGSOFT International
  Symposium on Software Testing and Analysis}, ser. ISSTA 2019.\hskip 1em plus
  0.5em minus 0.4em\relax ACM, 2019, pp. 19--30.

\bibitem{zhu2021recoder}
Q.~Zhu, Z.~Sun, Y.-a. Xiao, W.~Zhang, K.~Yuan, Y.~Xiong, and L.~Zhang, ``A
  syntax-guided edit decoder for neural program repair,'' in \emph{Proceedings
  of the 29th ACM Joint Meeting on European Software Engineering Conference and
  Symposium on the Foundations of Software Engineering}.\hskip 1em plus 0.5em
  minus 0.4em\relax New York, NY, USA: ACM, 2021, p. 341–353.

\bibitem{jiang2021cure}
N.~Jiang, T.~Lutellier, and L.~Tan, ``Cure: Code-aware neural machine
  translation for automatic program repair,'' \emph{2021 IEEE/ACM 43rd
  International Conference on Software Engineering (ICSE)}, May 2021.

\bibitem{ye2022rewardrepair}
H.~Ye, M.~Martinez, and M.~Monperrus, ``Neural program repair with
  execution-based backpropagation,'' in \emph{2022 IEEE/ACM 44th International
  Conference on Software Engineering (ICSE)}, 2022, pp. 1506--1518.

\bibitem{lutellier2020coconut}
T.~Lutellier, H.~V. Pham, L.~Pang, Y.~Li, M.~Wei, and L.~Tan, ``Coconut:
  Combining context-aware neural translation models using ensemble for program
  repair,'' in \emph{Proceedings of the 29th ACM SIGSOFT International
  Symposium on Software Testing and Analysis}, ser. ISSTA 2020.\hskip 1em plus
  0.5em minus 0.4em\relax New York, NY, USA: Association for Computing
  Machinery, 2020, p. 101–114.

\bibitem{sutskever2014mt}
I.~Sutskever, O.~Vinyals, and Q.~V. Le, ``Sequence to sequence learning with
  neural networks,'' 2014, arXiv:1409.3215.

\bibitem{xia2022alpharepair}
C.~S. Xia and L.~Zhang, ``Less training, more repairing please: Revisiting
  automated program repair via zero-shot learning,'' 2022, arXiv:2207.08281.

\bibitem{brown2020gpt3}
T.~B. Brown, B.~Mann, N.~Ryder, M.~Subbiah, J.~Kaplan, P.~Dhariwal,
  A.~Neelakantan, P.~Shyam, G.~Sastry, A.~Askell, S.~Agarwal, A.~Herbert-Voss,
  G.~Krueger, T.~Henighan, R.~Child, A.~Ramesh, D.~M. Ziegler, J.~Wu,
  C.~Winter, C.~Hesse, M.~Chen, E.~Sigler, M.~Litwin, S.~Gray, B.~Chess,
  J.~Clark, C.~Berner, S.~McCandlish, A.~Radford, I.~Sutskever, and D.~Amodei,
  ``Language models are few-shot learners,'' 2020, arXiv:2005.14165.

\bibitem{chen2021codex}
M.~Chen, J.~Tworek, H.~Jun, Q.~Yuan, H.~P. de~Oliveira~Pinto, J.~Kaplan,
  H.~Edwards, Y.~Burda, N.~Joseph, G.~Brockman, A.~Ray, R.~Puri, G.~Krueger,
  M.~Petrov, H.~Khlaaf, G.~Sastry, P.~Mishkin, B.~Chan, S.~Gray, N.~Ryder,
  M.~Pavlov, A.~Power, L.~Kaiser, M.~Bavarian, C.~Winter, P.~Tillet, F.~P.
  Such, D.~Cummings, M.~Plappert, F.~Chantzis, E.~Barnes, A.~Herbert-Voss,
  W.~H. Guss, A.~Nichol, A.~Paino, N.~Tezak, J.~Tang, I.~Babuschkin, S.~Balaji,
  S.~Jain, W.~Saunders, C.~Hesse, A.~N. Carr, J.~Leike, J.~Achiam, V.~Misra,
  E.~Morikawa, A.~Radford, M.~Knight, M.~Brundage, M.~Murati, K.~Mayer,
  P.~Welinder, B.~McGrew, D.~Amodei, S.~McCandlish, I.~Sutskever, and
  W.~Zaremba, ``Evaluating large language models trained on code,'' 2021,
  arXiv:2107.03374.

\bibitem{feng2020codebert}
Z.~Feng, D.~Guo, D.~Tang, N.~Duan, X.~Feng, M.~Gong, L.~Shou, B.~Qin, T.~Liu,
  D.~Jiang, and M.~Zhou, ``Codebert: A pre-trained model for programming and
  natural languages,'' 2020, arXiv:2002.08155.

\bibitem{kolak2022patch}
S.~D. Kolak, R.~Martins, C.~L. Goues, and V.~J. Hellendoorn, ``Patch generation
  with language models: Feasibility and scaling behavior,'' in \emph{Deep
  Learning for Code Workshop}, 2022.

\bibitem{prenner2021codexws}
J.~A. Prenner, H.~Babii, and R.~Robbes, ``Can openai's codex fix bugs?: An
  evaluation on quixbugs,'' in \emph{2022 IEEE/ACM International Workshop on
  Automated Program Repair (APR)}, 2022, pp. 69--75.

\bibitem{fried2022incoder}
D.~Fried, A.~Aghajanyan, J.~Lin, S.~Wang, E.~Wallace, F.~Shi, R.~Zhong, W.-t.
  Yih, L.~Zettlemoyer, and M.~Lewis, ``Incoder: A generative model for code
  infilling and synthesis,'' 2022, arXiv:2204.05999.

\bibitem{papineni2002bleu}
K.~Papineni, S.~Roukos, T.~Ward, and W.-J. Zhu, ``Bleu: A method for automatic
  evaluation of machine translation,'' in \emph{Proceedings of the 40th Annual
  Meeting on Association for Computational Linguistics}, ser. ACL '02, 2002, p.
  311–318.

\bibitem{liu2019finetune}
Y.~Liu, ``Fine-tune bert for extractive summarization,'' 2019,
  arXiv:1903.10318.

\bibitem{yang2020xlnet}
Z.~Yang, Z.~Dai, Y.~Yang, J.~Carbonell, R.~Salakhutdinov, and Q.~V. Le,
  ``Xlnet: Generalized autoregressive pretraining for language understanding,''
  2020, arXiv:1906.08237.

\bibitem{vaswani2017attention}
A.~Vaswani, N.~Shazeer, N.~Parmar, J.~Uszkoreit, L.~Jones, A.~N. Gomez,
  L.~Kaiser, and I.~Polosukhin, ``Attention is all you need,'' 2017,
  arXiv:1706.03762.

\bibitem{reynolds2021prompt}
L.~Reynolds and K.~McDonell, ``Prompt programming for large language models:
  Beyond the few-shot paradigm,'' 2021, arXiv:2102.07350.

\bibitem{kaplan2020scaling}
J.~Kaplan, S.~McCandlish, T.~Henighan, T.~B. Brown, B.~Chess, R.~Child,
  S.~Gray, A.~Radford, J.~Wu, and D.~Amodei, ``Scaling laws for neural language
  models,'' 2020, arXiv:2001.08361.

\bibitem{devlin2018bert}
J.~Devlin, M.-W. Chang, K.~Lee, and K.~Toutanova, ``Bert: Pre-training of deep
  bidirectional transformers for language understanding,'' 2018,
  arXiv:1810.04805.

\bibitem{gpt-neo}
\BIBentryALTinterwordspacing
S.~Black, L.~Gao, P.~Wang, C.~Leahy, and S.~Biderman, ``{GPT-Neo: Large Scale
  Autoregressive Language Modeling with Mesh-Tensorflow},'' Mar. 2021.
  [Online]. Available: \url{https://doi.org/10.5281/zenodo.5297715}
\BIBentrySTDinterwordspacing

\bibitem{raffel2020t5}
C.~Raffel, N.~Shazeer, A.~Roberts, K.~Lee, S.~Narang, M.~Matena, Y.~Zhou,
  W.~Li, and P.~J. Liu, ``Exploring the limits of transfer learning with a
  unified text-to-text transformer,'' \emph{J. Mach. Learn. Res.}, jan 2020.

\bibitem{lewis2019bart}
M.~Lewis, Y.~Liu, N.~Goyal, M.~Ghazvininejad, A.~Mohamed, O.~Levy, V.~Stoyanov,
  and L.~Zettlemoyer, ``Bart: Denoising sequence-to-sequence pre-training for
  natural language generation, translation, and comprehension,'' 2019,
  arXiv:1910.13461.

\bibitem{aghajanyan2022cm3}
A.~Aghajanyan, B.~Huang, C.~Ross, V.~Karpukhin, H.~Xu, N.~Goyal, D.~Okhonko,
  M.~Joshi, G.~Ghosh, M.~Lewis, and L.~Zettlemoyer, ``Cm3: A causal masked
  multimodal model of the internet,'' 2022, arXiv:2201.07520.

\bibitem{hindle2012softwarenatural}
A.~Hindle, E.~T. Barr, Z.~Su, M.~Gabel, and P.~Devanbu, ``On the naturalness of
  software,'' in \emph{Proceedings of the 34th International Conference on
  Software Engineering}, ser. ICSE '12, 2012, p. 837–847.

\bibitem{HuggingFaceWebPage}
``Hugging face,'' 2022, \url{https://huggingface.co}.

\bibitem{gpt-j}
B.~Wang and A.~Komatsuzaki, ``{GPT-J-6B: A 6 Billion Parameter Autoregressive
  Language Model},'' \url{https://github.com/kingoflolz/mesh-transformer-jax},
  May 2021.

\bibitem{gpt-neox-20b}
S.~Black, S.~Biderman, E.~Hallahan, Q.~Anthony, L.~Gao, L.~Golding, H.~He,
  C.~Leahy, K.~McDonell, J.~Phang, M.~Pieler, U.~S. Prashanth, S.~Purohit,
  L.~Reynolds, J.~Tow, B.~Wang, and S.~Weinbach, ``{GPT-NeoX-20B}: An
  open-source autoregressive language model,'' in \emph{Proceedings of the ACL
  Workshop on Challenges \& Perspectives in Creating Large Language Models},
  2022, arXiv:2204.06745.

\bibitem{gao2020pile}
L.~Gao, S.~Biderman, S.~Black, L.~Golding, T.~Hoppe, C.~Foster, J.~Phang,
  H.~He, A.~Thite, N.~Nabeshima \emph{et~al.}, ``The pile: An 800gb dataset of
  diverse text for language modeling,'' 2020, arXiv:2101.00027.

\bibitem{wang2021codet5}
S.~J. Yue~Wang, Weishi~Wang and S.~C. Hoi, ``Codet5: Identifier-aware unified
  pre-trained encoder-decoder models for code understanding and generation,''
  in \emph{Proceedings of the 2021 Conference on Empirical Methods in Natural
  Language Processing, EMNLP 2021}, 2021.

\bibitem{husain2020codesearchnet}
H.~Husain, H.-H. Wu, T.~Gazit, M.~Allamanis, and M.~Brockschmidt,
  ``Codesearchnet challenge: Evaluating the state of semantic code search,''
  2020, arXiv:1909.09436.

\bibitem{BigQuery}
``Bigquery github repos,'' 2022,
  \url{https://console.cloud.google.com/marketplace/details/github/github-repos}.

\bibitem{CodexSuffixAPI}
``Codex suffix api,''
  \url{https://beta.openai.com/docs/api-reference/completions/create\#completions/create-suffix},
  2022.

\bibitem{abreu2007ochiai}
R.~Abreu, P.~Zoeteweij, and A.~J. van Gemund, ``On the accuracy of
  spectrum-based fault localization,'' in \emph{Testing: Academic and
  Industrial Conference Practice and Research Techniques - MUTATION
  (TAICPART-MUTATION 2007)}, 2007, pp. 89--98.

\bibitem{zhang2013injecting}
L.~Zhang, L.~Zhang, and S.~Khurshid, ``Injecting mechanical faults to localize
  developer faults for evolving software,'' \emph{ACM SIGPLAN Notices},
  vol.~48, no.~10, pp. 765--784, 2013.

\bibitem{holtzman2019nucleus}
A.~Holtzman, J.~Buys, L.~Du, M.~Forbes, and Y.~Choi, ``The curious case of
  neural text degeneration,'' 2019, arXiv:1904.09751.

\bibitem{wang2022pcc}
Y.~Wang, J.~Yang, Y.~Lou, M.~Wen, and L.~Zhang, ``Attention: Not just another
  dataset for patch-correctness checking,'' 2022, arXiv:2207.06590.

\bibitem{PyTorchWebPage}
``Pytorch,'' 2022, \url{http://pytorch.org}.

\bibitem{OpenAI}
``Openai api,'' 2022, \url{https://openai.com/api}.

\bibitem{li2022alphacode}
Y.~Li, D.~Choi, J.~Chung, N.~Kushman, J.~Schrittwieser, R.~Leblond, T.~Eccles,
  J.~Keeling, F.~Gimeno, A.~D. Lago, T.~Hubert, P.~Choy, C.~d.~M. d'Autume,
  I.~Babuschkin, X.~Chen, P.-S. Huang, J.~Welbl, S.~Gowal, A.~Cherepanov,
  J.~Molloy, D.~J. Mankowitz, E.~S. Robson, P.~Kohli, N.~de~Freitas,
  K.~Kavukcuoglu, and O.~Vinyals, ``Competition-level code generation with
  alphacode,'' 2022, arXiv:2203.07814.

\bibitem{just2014dfj}
R.~Just, D.~Jalali, and M.~D. Ernst, ``Defects4j: A database of existing faults
  to enable controlled testing studies for java programs,'' ser. ISSTA
  2014.\hskip 1em plus 0.5em minus 0.4em\relax New York, NY, USA: Association
  for Computing Machinery, 2014, p. 437–440.

\bibitem{lin2017quixbug}
D.~Lin, J.~Koppel, A.~Chen, and A.~Solar-Lezama, ``Quixbugs: A multi-lingual
  program repair benchmark set based on the quixey challenge,'' ser. SPLASH
  Companion 2017.\hskip 1em plus 0.5em minus 0.4em\relax New York, NY, USA:
  Association for Computing Machinery, 2017, p. 55–56.

\bibitem{legoues2015manybugs}
C.~Le~Goues, N.~Holtschulte, E.~K. Smith, Y.~Brun, P.~Devanbu, S.~Forrest, and
  W.~Weimer, ``The manybugs and introclass benchmarks for automated repair of c
  programs,'' \emph{IEEE Transactions on Software Engineering}, vol.~41,
  no.~12, pp. 1236--1256, 2015.

\bibitem{drain2021deepdebug}
D.~Drain, C.~B. Clement, G.~Serrato, and N.~Sundaresan, ``Deepdebug: Fixing
  python bugs using stack traces, backtranslation, and code skeletons,'' 2021,
  arXiv:2105.09352.

\bibitem{li2020dlfix}
Y.~Li, S.~Wang, and T.~N. Nguyen, ``Dlfix: Context-based code transformation
  learning for automated program repair,'' in \emph{Proceedings of the ACM/IEEE
  42nd International Conference on Software Engineering}, ser. ICSE '20.\hskip
  1em plus 0.5em minus 0.4em\relax New York, NY, USA: Association for Computing
  Machinery, 2020, p. 602–614.

\bibitem{chen2018sequencer}
Z.~Chen, S.~Kommrusch, M.~Tufano, L.-N. Pouchet, D.~Poshyvanyk, and
  M.~Monperrus, ``Sequencer: Sequence-to-sequence learning for end-to-end
  program repair,'' \emph{IEEE Transaction on Software Engineering}, 2019.

\bibitem{jiang2018simfix}
J.~Jiang, Y.~Xiong, H.~Zhang, Q.~Gao, and X.~Chen, ``Shaping program repair
  space with existing patches and similar code,'' in \emph{Proceedings of the
  27th {ACM} {SIGSOFT} International Symposium on Software Testing and
  Analysis, {ISSTA} 2018, Amsterdam, The Netherlands, July 16-21, 2018}, F.~Tip
  and E.~Bodden, Eds.\hskip 1em plus 0.5em minus 0.4em\relax {ACM}, 2018, pp.
  298--309.

\bibitem{chen2017jaid}
L.~Chen, Y.~Pei, and C.~A. Furia, ``Contract-based program repair without the
  contracts,'' in \emph{2017 32nd IEEE/ACM International Conference on
  Automated Software Engineering (ASE)}, 2017, pp. 637--647.

\bibitem{martinez2015automatic}
M.~Martinez, T.~Durieux, J.~Xuan, R.~Sommerard, and M.~Monperrus, ``Automatic
  repair of real bugs: An experience report on the defects4j dataset,'' 2015,
  arXiv:1505.07002.

\bibitem{tufano2018empstudy}
M.~Tufano, C.~Watson, G.~Bavota, M.~Di~Penta, M.~White, and D.~Poshyvanyk, ``An
  empirical investigation into learning bug-fixing patches in the wild via
  neural machine translation,'' in \emph{Proceedings of the 33rd ACM/IEEE
  International Conference on Automated Software Engineering}, 2018, p.
  832–837.

\bibitem{pilefix1}
``jetbrick-template-2x object comparison code,'' 2022,
  \url{https://github.com/subchen/jetbrick-template-2x/blob/def3107e2878aa5bee32ac2ba3be8e241fba4a64/src/main/java/jetbrick/template/parser/ast/ALU.java#L421-L448}.

\bibitem{pilefix2}
``goclipse object comparison code,'' 2022,
  \url{https://github.com/GoClipse/goclipse/blob/e135d3a69e6498e278521c2542cee3808bd1377d/plugin_tooling/src-util/melnorme/utilbox/core/CoreUtil.java#L28-L30}.

\bibitem{pilefix3}
``teiid object comparison code,'' 2022,
  \url{https://github.com/teiid/teiid/blob/21c93a6fd4be2528f95224f99905d74479862d1b/federate-common-core/src/main/java/com/metamatrix/core/util/EquivalenceUtil.java#L49-L57}.

\bibitem{pilefix4}
``Groza object comparison code,'' 2022,
  \url{https://github.com/IoT-Technology/Groza/blob/fbafceef53d646025046990ffbd89bf701c56b45/dao/src/main/java/com/sanshengshui/server/dao/util/mapping/JsonTypeDescriptor.java#L49-L58}.

\bibitem{asad2019impact}
M.~Asad, K.~K. Ganguly, and K.~Sakib, ``Impact analysis of syntactic and
  semantic similarities on patch prioritization in automated program repair,''
  in \emph{2019 IEEE International Conference on Software Maintenance and
  Evolution (ICSME)}, 2019, pp. 328--332.

\bibitem{xin2017leveraging}
Q.~Xin and S.~P. Reiss, ``Leveraging syntax-related code for automated program
  repair,'' in \emph{2017 32nd IEEE/ACM International Conference on Automated
  Software Engineering (ASE)}.\hskip 1em plus 0.5em minus 0.4em\relax IEEE,
  2017, pp. 660--670.

\bibitem{correctPatchandDataset}
``Dataset,'' 2022, \url{https://figshare.com/s/temp}.

\end{thebibliography}

\end{document}